# Dynamic Behavior of Tandem Perforated Elastic Vortex Generators Using Two-Way Coupled Fluid-Structure Interaction Simulations

Karan Kakroo[a], Hamid Sadat[a,1]

[a] Department of Mechanical Engineering, University of North Texas, Denton, Texas, USA

## Abstract

This study presents high-fidelity, two-way coupled fluid–structure interaction (FSI) simulations to investigate the dynamic behavior of tandem perforated elastic vortex generators (PEVGs) across a wide range of bending rigidity, mass ratio, and porosity, at a fixed Reynolds number and interspacing. Comparative simulations with non-perforated EVGs are also performed. Three response modes—Lodging, Vortex-Induced Vibration (VIV), and Static Reconfiguration—are observed in both configurations, while a distinct Cavity Oscillation mode emerges exclusively in non-perforated tandem EVGs. This mode is entirely suppressed with porosity due to disruption of the low-pressure cavity and increased flow transmission through pores. Frequency analyses reveal that VIV is consistently locked onto the second natural frequency, whereas the Cavity Oscillation mode is locked onto the first natural frequency and closely aligns with the first Rossiter mode, underscoring its distinct physical origin. Perforation modifies the natural frequency of the EVGs, shifting the lock-in and mode transitions toward lower bending rigidity and higher mass ratio values, and reducing oscillation amplitudes due to motion damping. Drag analysis shows consistently higher upstream drag due to wake shielding, while porosity reduces upstream drag and increases downstream drag by restoring streamwise momentum. Flow visualizations demonstrate that vortex shedding originates at the EVG tips, with perforated configurations producing smaller, more dissipative vortical structures. These results establish that porosity fundamentally alters dynamic regimes, suppresses cavity-driven instabilities, and enables passive modulation of wake dynamics in tandem EVG systems.

## Nomenclature

| Symbol | Description | | |
|---|---|---|---|
| L | Length of the EVG | p | Pressure |
| Δ | Thickness of the EVG | μ | Dynamic viscosity |
| w | Depth of the EVG | S | Strain-rate tensor |
| D | Distance between upstream and downstream EVGs | $\varepsilon$ | Green-Lagrange strain tensor |
| N | Number of pores on the EVG | E | Young's modulus |
| a | Side length of square perforations | ν | Poisson's ratio |
| b | Spacing between adjacent perforations | λ | Lame constant |
| ϕ | Porosity of the EVG | γ | Dimensionless bending rigidity |
| v | Velocity vector | β | Dimensionless mass ratio |
| $v^m$ | Grid velocity vector | $\rho_f$ | Fluid density |
| ρ | Mass density | $\rho_S$ | Solid (EVG) density |
| σ | Cauchy stress tensor | I | Second moment of area |
| F | Body force | $F_P$ | Pressure loading acting over the surface |
| | | U | Free-stream velocity |
| | | Re | Reynolds number |

*Corresponding author
Email address: hamid.sadat@unt.edu

| $n_f$ | Number of EVGs | | $f_{ni}$ | Non-dimensional $i^{th}$ natural frequency |
|---|---|---|---|---|
| $\Delta t$ | Non-dimensional time | | $k_1, k_2$ | Constants for first and second natural frequencies respectively |
| $t$ | Dimensional time | | | |
| $\theta$ | Inclination angle of EVG | | $C_m$ | Added mass coefficient |
| $\theta_{ave}$ | Time-averaged inclination angle | | $X$ | Porosity-dependent mass correction factor |
| $\theta_H$ | Oscillatory amplitude of inclination angle | | $f_c$ | Rossiter-mode dimensionless oscillation frequency |
| $\theta_U$ | Inclination angle for upstream EVG | | | |
| $\theta_D$ | Inclination angle for downstream EVG | | $n$ | Rossiter mode number |
| $\theta_S$ | Inclination angle for single EVG | | $\kappa$ | Empirical constant in Rossiter's formula |
| $\theta^{NP}$ | Inclination angle for non-perforated EVG | | $\alpha$ | Empirical constant in Rossiter's formula |
| $\theta^P$ | Inclination angle for perforated EVG | | $M$ | Mach number |
| $EI_{NP}$ | Bending stiffness for non-perforated EVG | | $C_D$ | Drag coefficient |
| $EI_P$ | Bending stiffness for perforated EVG | | $\omega$ | Vorticity |
| $K$ | Correction factor for bending stiffness due to perforation | | | |

## 1. Introduction

The interaction between fluid flow and flexible structures is ubiquitous in both nature and engineering sectors, often categorized as a fluid-structure interaction (FSI) problem. Over the years, many researchers and engineers have increasingly applied the insights from FSI studies to the design and optimization of vortex generators (VGs), as they play an important role in intensifying mixing processes and enhancing heat transfer efficiency[1-5] across a diverse array of industrial and engineering sectors[6-13]. VGs have been extensively studied for their ability to induce swirling motion and introduce additional turbulence into the flow, thereby substantially improving and enhancing the mixing and heat transfer[14-20]. Studies have spanned both rigid (RVGs) [21-30] and flexible or elastic vortex generators (EVGs)[31-47], with findings concluding that EVGs significantly increase mixing and heat transfer while incurring a lower pressure drop compared to their rigid counterparts[31,33,34,38,44,45,48].

VGs and EVGs have demonstrated utility across a wide Reynolds number spectrum in both microscale and macroscale systems. At low Reynolds numbers, they have been used in tissue engineering applications, such as perfusion bioreactors and organ-on-chip platforms, to improve nutrient mixing and cell viability[49-51]. In electronics cooling, flexible VGs enable dynamic redirection of coolant flow in compact channels to mitigate thermal hotspots[52-55]. In microfluidic systems such as lab-on-a-chip devices, vortex generators have been effectively employed to enhance mixing efficiency at low Reynolds numbers, enabling uniform distribution of reagents for biomedical diagnostics, chemical synthesis, and organ-on-chip platforms[56-58]. In biomedical engineering, VGs are extensively used to reduce the risk of heart disease (hemolysis and platelet activation) via reduction of turbulent stresses, by placing them at the surface of the valve leaflets in order to delay and mitigate flow separations by bringing momentum from the main-stream into the boundary layer, which manipulate the flow to restrict detrimental effects or promote the beneficial ones[59-62]. Meanwhile, at high Reynolds numbers, VGs are widely implemented in aerospace engineering for shock induced separation control, drag reduction and boundary layer control on hypersonic vehicles, as well as in thermal energy systems such as compact heat exchangers and thermal batteries[9,29,63-68]. However, majority of these studies have been limited to single EVG configurations.

Very limited studies have employed multiple or tandem EVGs [32,34,35,37,38,41,42,69–82] to investigate the enhancement in heat transfer and mixing due to the interaction of EVGs with each other and with the incoming flow in tandem configurations. These studies have demonstrated that using tandem configurations, the vortex-vortex interaction significantly improves and enhances the mixing and heat transfer efficiency compared to the rate at which it is increased using a single EVG. Additionally, these studies have revealed that the oscillation of upstream EVG significantly modulates/alters the oscillation of downstream EVG, because of which the vortices shedding from both the upstream and downstream EVGs interact with each other, hence causing an increased turbulence within the fluid flow, thereby enhancing heat transfer and mixing at a rate which is beyond what is achievable with a single EVG. Nevertheless, despite their promising results, existing studies on tandem EVGs have predominantly focused on non-perforated structures, which often comes with a penalty of increase in pressure drop.

To mitigate this, perforated EVGs (PEVGs) have been proposed to reduce pressure drop while maintaining or enhancing the mixing and heat transfer characteristics even further. Previous studies on perforated flexible structures [45,83–106] have highlighted the benefits of introducing perforations and have revealed that the pressure drop is often less for perforated structures compared to their non-perforated counterparts. Perforations allow for flow bleeding, which reduces pressure differentials and results in more dissipative, less coherent vortex structures. In our recent study [45], we extended this concept by conducting a detailed investigation on a single wall-mounted perforated elastic vortex generator (PEVG) using high-fidelity two-way coupled FSI simulations. That study revealed that single PEVGs exhibit three distinct dynamic response modes—Lodging, Vortex-Induced Vibration (VIV), and Static Reconfiguration—depending on parameters such as bending rigidity, mass ratio, Reynolds number, and porosity. A key finding was that porosity reduces oscillation amplitudes, alters transition boundaries between regimes, and often decreases drag/pressure drop. Flow field analysis also showed that perforations induce bleeding flow which stabilizes the wake and generates smaller vortical structures compared to the non-perforated EVG. Despite these insights for single PEVGs, the dynamic behavior and wake interactions in tandem configurations remain largely unexplored. In particular, no systematic study has addressed how perforation modifies the coupled dynamics, flow-induced forces, and oscillation modes of tandem PEVGs. This gap motivates the present investigation.

In the present study, we investigated two tandem perforated EVGs (PEVGs) using high-fidelity 2-way coupled Fluid Structure Interaction (FSI) simulations, to analyze the dynamic response of upstream and downstream EVGs over a wide range of dimensionless parameters, including bending rigidity, mass ratio, and porosity at a fixed Reynolds number and fixed interspace between two EVGs. Additional simulations for non-perforated EVGs are conducted for comparison. The response characteristics such as inclination angle; phase portrait; and response harmonics, as well as the local flow are analyzed to understand the role of perforation on fluid and solid interactions.

## 2. Geometry Information

The geometric configuration of the tandem elastic vortex generators (EVGs) considered in this study is depicted by Figure 1. Each EVG is composed of an elastic material with dimensions of L = 0.1 m (length), δ = 0.008 m (thickness), and w = 0.06 m (depth). The geometric configuration, including the EVG dimensions and aspect ratio, was adopted from the closest relevant benchmark in the literature, Zhang et al.[82], which investigated the behavior of non-perforated tandem EVGs. To avoid artificial stress concentrations and better reflect realistic structural geometries, the top edges of both upstream and downstream EVGs are rounded with a corner radius of 0.004 m (0.04 L). In the tandem arrangement, the upstream and downstream EVGs are closely spaced D=1L apart from each other, to examine their dynamic response due to wake interactions and shielding effects. To investigate the influence of porosity, a perforated configuration incorporating N=15 uniformly distributed square pores was considered, arranged in a 5 × 3 matrix along the EVG's length and depth respectively, as depicted by Figure 1b. Different porosity levels $\phi = \frac{Na^2}{Lw} = [0, 0.1, 0.25, \text{ and } 0.5]$) were studied by adjusting the size of each pore a. The associated side lengths of the square perforations were set as a= 0, 0.006325, 0.01, and 0.0141421 m respectively, while the spacing between adjacent perforations varied as b = 0.02, 0.013675, 0.01 and 0.0058579 m respectively, corresponding to different porosity levels. This systematic variation enables a comprehensive analysis of porosity effects on both upstream and downstream EVGs, particularly regarding wake interactions, and EVGs dynamic response.

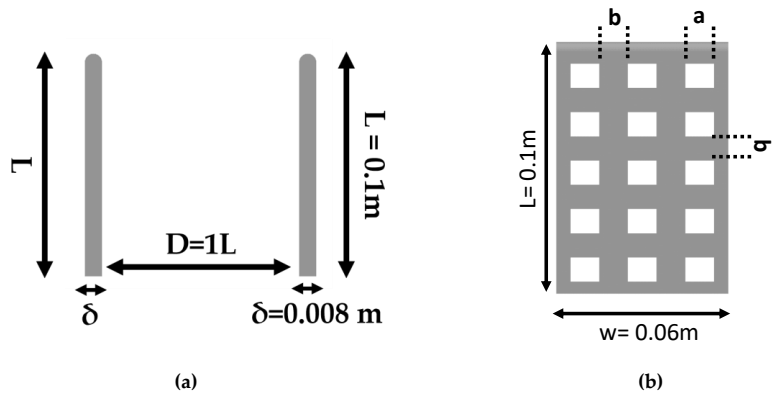

**Figure 1: Geometric configuration of Tandem EVGs (a) Front view; and (b) Side view.**

### 3. Computational Domain, Grid Information, and Boundary Conditions

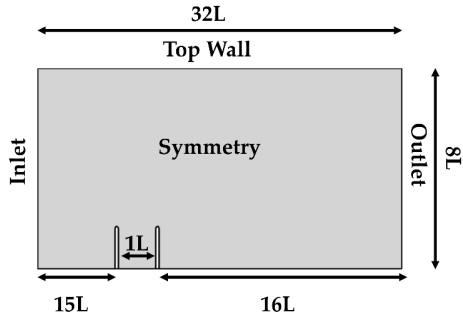

**Figure 2: Computation domain.**

The computation domain for this study is a rectangular box extending from [0, 32L] in the streamwise direction and [0, 8L] in the transverse direction, as depicted by Figure 2. The total streamwise length is 32L with the inlet positioned 15L upstream of leading edge of the upstream EVG, and the outlet located 16L downstream of the trailing edge of the downstream EVG. Both EVGs are rigidly clamped at their bases and fixed to the bottom wall of the computational domain.

Unstructured grids for both the solid and fluid domains are employed, as depicted by Figure 3a and Figure 3b respectively.

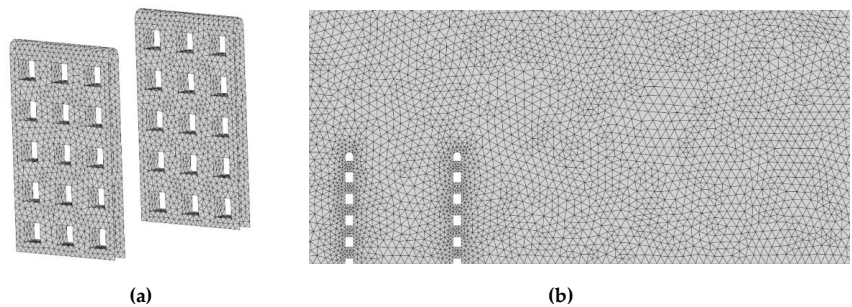

**Figure 3: Grid information for the tandem EVG configuration: (a) solid domain mesh; and (b) fluid domain mesh.**

For the solid domain, the mesh size was assigned as 0.060166L far from pores and refined to 0.02L near pores. An additional targeted refinement within the porous regions was employed to accurately capture the local flow physics. The resulting mesh resolution varied with different porosity levels, yielding approximately 54-72 cells across the depth, 98-144 cells along the length, and 9 cells along the thickness of EVG. This yielded a final mesh comprising approximately 26K-35K nodes, and 76K-101K elements, providing sufficient resolution to capture the solid behavior, as shown by Figure 3a.

The fluid domain was also discretized using an unstructured mesh with a size of 0.072288L at far field. To accurately capture and resolve complex physics involving the generation of vortical structures in the near wake region, a finer mesh was employed on the FSI-coupled surfaces as well as near the pores, utilizing a face sizing of 0.03L. Along the streamwise direction, the grid consisted of 447 cells, distributed as 209 cells from inlet to the upstream EVG, 16 cells within the inter-spacing (D = 1L) of two tandem EVGs, and 222 cells extending from downstream EVG to the outlet, ensuring adequate resolution for wake development and interaction. Along the transverse direction, the fluid domain included 111 cells; while in the depth direction, the resolution varied from 10 cells in regions far from the EVG to approximately 21-25 cells in the proximity of the EVG, depending upon the porosity level being investigated. This strategy yielded a total of 2M unstructured cells for the fluid domain. Additionally, a grid independence study was conducted to verify and ensure that the results were not influenced by the grid resolution. A systematic refinement approach was applied to both the solid and fluid domains using a refinement factor of $\sqrt{2}$, yielding three grid sets: G1 (finest), G2 (intermediate), and G3 (coarsest).

The boundary conditions used in this study are defined in Table 1.

**Table 1: Boundary conditions**

| Boundary Condition | Type | |
|---|---|---|
| | U | P |
| **Fluid component** | | |
| Inlet | $u = \vec{U}_{inlet}$ | $\frac{dP}{dn} = 0$ |
| Outlet | $\frac{d\vec{U}}{dn} = 0$ | P=P$_{constant}$ |
| Bottom wall | $\vec{U}$=0 | $\frac{dP}{dn} = 0$ |
| Top wall | $\vec{U}. \text{n}=0$ | $\frac{dP}{dn} = 0$ |
| Front and back | $\frac{d\vec{U}}{dn} = 0$ | $\frac{dP}{dn} = 0$ |
| **Solid component** | | |
| FSI Coupled faces | $\vec{U}_{fluid} = \vec{U}_{solid}$ | P$_{fluid}$ = P$_{solid}$ |
| Bottom face (clamped) | $\vec{U}$=0 | $\frac{dP}{dn} = 0$ |

The fluid domain is subjected to a uniform velocity inlet boundary condition, with ( $u = \vec{U}_{inlet}$ ), corresponding to a fixed Reynolds number of Re = 400. A zero normal pressure gradient is applied at the inlet ($\frac{dP}{dn} = 0$). At the outlet, an outflow boundary condition is specified, with a zero-gradient condition for velocity ($\frac{d\vec{U}}{dn} = 0$) along with a fixed pressure (P=P$_{constant}$). The top boundary is treated as a slip wall, while the bottom wall is modeled as a no-slip surface. Symmetry boundary conditions are enforced along the spanwise side boundaries (front and back), where both ($\frac{d\vec{U}}{dn} = 0$) and ($\frac{dP}{dn} = 0$) are satisfied. In the solid domain, the bottom surfaces of both upstream and downstream PEVGs are fully clamped, enforcing a fixed vertical orientation throughout the simulations. At the fluid-structure interfaces, continuity of velocity and normal stress is enforced: $(\overline{U}_{fluid} = \overline{U}_{solid})$ and $(P_{fluid} = P_{solid})$, ensuring fully coupled fluid–structure interaction throughout the simulation.

### 4. Fluid and Structure solvers

A partitioned two-way fluid-structure interaction (FSI) coupling scheme is utilized for all the simulations conducted in this study, where the governing equations for the fluid and solid components are solved iteratively. The governing equations are given as follows:

$$\frac{\partial \rho}{\partial t} + \nabla \cdot [\rho(v - v^m)] = 0 \tag{1}$$

$$\rho \frac{\partial v}{\partial t} + \rho[(v - v^m) \cdot \nabla]v = \nabla \cdot \sigma + \rho F \tag{2}$$

where $\rho$, $v$, $v^m$, $\sigma$, and F represents the mass density, particle velocity, grid velocity, Cauchy stress tensor, and body force, respectively. The Cauchy stress tensor for the fluid component is defined as $(\sigma = -pI + 2\mu S)$, where p is the thermodynamic pressure, $\mu$ is the dynamic viscosity, I is the second rank unity tensor and S is the strain-rate tensor. The Cauchy stress tensor for a linear elastic solid is defined as $(\sigma = 2\mu\varepsilon + \lambda\nabla \cdot UI)$, where $\mu$=E/2(1+$\nu$) and $\lambda$= $\nu$E/(1+ $\nu$)(1-2$\nu$) are Lame's constants, $\varepsilon$ is the Green-Lagrange strain tensor, E is the Young's modulus, and $\nu$ is poison's ratio.

The simulations were performed using the ANSYS software package[107,108], where ANSYS Fluent serves as the fluid solver and ANSYS Transient Structural is employed for the structural domain. These solvers are coupled via ANSYS System Coupling, which facilitates an implicit two-way exchange of force and displacement across the fluid–structure interface at every time step. For the fluid component, the governing equations are solved using the finite volume method. Pressure–velocity coupling is handled using a coupled algorithm to ensure simultaneous satisfaction of continuity and momentum equations. Spatial discretization is performed using a second-order upwind scheme for the convective terms and the least-squares cell-based method for gradient terms. Time integration is carried out using a second-order implicit formulation, which enhances stability and accuracy for capturing unsteady, oscillatory flow phenomena common in FSI problems. Mesh motion is handled using a spring-based analogy[109–112], where all point-to-point mesh connections are replaced with springs. If the mesh quality deteriorates due to excessive skewness or deformation, local grid regeneration is triggered via a dynamic remeshing technique. Otherwise, mesh smoothing is applied to maintain numerical stability and accuracy without altering the topology.

On the structural side, ANSYS Transient Structural employs an implicit time integration scheme based on the Newmark-beta method to resolve the dynamics of the flexible EVGs[107,108]. The structures are modeled as linearly elastic, and no geometric or material nonlinearities are considered. Interface continuity is enforced at each time step by exchanging pressure and shear force data from the fluid domain to the structure and updating the fluid mesh using the resulting displacements from the structural solver. This implicit coupling ensures numerical convergence and stability throughout the simulation.

## 5. Test Conditions

The governing equation for EVG motion, shown in Equation (2) can be simplified using Euler-Bernoulli beam theory under the assumption of negligible shear stress resultant, leading to the following dimensionless equation:

$$(\gamma)\frac{\partial^4 X^*}{\partial s^{*4}} + (\beta)\frac{\partial^2 X^*}{\partial t^{*2}} = \frac{F_p}{\rho_f U^2 L^2} \tag{3}$$

where $\gamma = \frac{EI}{\rho_f U^2_\infty L^3}$ and $\beta = \frac{\rho_s \delta}{\rho_f L}$ represent two dimensionless terms namely bending rigidity and mass ratio respectively. Here, $\rho_f$ denotes the fluid density, $\rho_s$ is the density of the EVG, E is the Young's modulus, I is the second moment of area, $\delta$ is the thickness of the EVG, and L is its length. The term $\frac{F_p}{\rho_f U^2 L^2}$ represents the dimensionless fluid force acting on the EVG, which is a function of Reynolds number, defined as $Re = \frac{\rho U L}{\mu}$, where $\mu$ is the dynamic viscosity of the fluid. Therefore, the dynamic response of EVGs in a fluid flow is governed by dimensionless parameters $\gamma$, $\beta$, and Re along with porosity levels $\phi$ that characterize their structural behavior and flow-induced interactions. In this study, Reynolds number is fixed at Re=400, while the dimensionless parameters considered in this study range from $0.001 \leq \gamma \leq 5$, and $0.1 \leq \beta \leq 4$, covering a broad range of bending rigidity/flexibility and mass ratio conditions. The use of these broad range of parameters also enabled meaningful comparisons with Zhang et al.'s[82] results, which we used for validation in both the single and tandem non-perforated EVG configurations, discussed in Section 6.4. Different porosity levels are considered ranging from $\phi$= [0, 0.1, 0.25, and 0.5], as discussed earlier. The test conditions are summarized in Table 2.

**Table 2: Summary of Test Conditions**

| $n_f$ | $\gamma$ | $\beta$ | Re | D |
|---|---|---|---|---|
| 2 | $10^{-3} - 5$ | $0.1 - 4$ | 400 | 1.0 L |

This study investigates the structural response, hydrodynamic loads, as well as local flow features such as vortical structures and velocity fields. To specifically assess the structural response of the EVGs, this study utilizes the inclination angle ($\theta$), previously introduced in our study[45] as a key parameter for evaluating both structural deflection/bending and oscillatory behavior of the EVG. The inclination angle is defined as the angle between the EVG's root-to-tip and the free-stream direction. The analysis focuses on two primary metrics: 1) the time-averaged inclination angle, representing the mean structural deformation over time, and 2) the oscillatory amplitude, capturing the extent of angular variations during oscillation. These are expressed as:

$$\theta \text{ave} = \frac{1}{t_1 - t_0}\int_{t_0}^{t_1}\theta(t)dt \tag{4}$$

$$\theta_H = \{\max[\theta(t)] - \min[\theta(t)]\}_{t_0 \leq t \leq t_1} \tag{5}$$

where the difference $t_1 - t_0$ denotes the oscillation period.

## 6. Results and discussions

### 6.1. Domain Size Verification

The computational domain extends from 0 to 32L in the streamwise direction, where L=0.1 m denotes the EVG length. The inlet is positioned 15L upstream of the leading edge of the upstream EVG, while the outlet is located 16L downstream of the trailing edge of the downstream EVG. To evaluate the adequacy of this domain size, two shorter domains were also tested: D2 with a total length of 21.3L, and D3 with 10.7L. The streamwise velocity predicted at a location 0.5L downstream of the leading edge of the upstream EVG was compared across all three domains—D1 (32L), D2 (21.3L), and D3 (10.7L). All simulation were performed for $\gamma = 5$, at a fixed $\beta = 1$ and Re = 400 condition for $\phi = 0$. The comparison, shown in Figure 4, reveals that the shortest domain D3 deviates notably in velocity trend and magnitude, particularly in the wake region. However, the velocity profiles for D2 and D1 are nearly identical, suggesting that extending the domain beyond D1 offers no additional improvement in capturing the flow features. Therefore, the domain D1 used for all simulations in this study accurately resolves the wake dynamics.

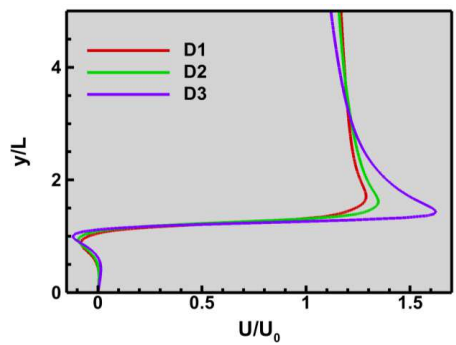

**Figure 4: Effect of domain size on the streamwise velocity profiles.**

### 6.2. Grid Verification

The grid verification simulations using grids G1, G2, and G3 were performed for $\gamma = 5$, at a fixed $\beta = 1$ and Re = 400 condition for $\phi = 0$. To evaluate the influence of grid resolution on flow predictions, the streamwise velocity profile was extracted at the midplane between the two tandem EVGs and compared across all grid levels, as shown by Figure 5.

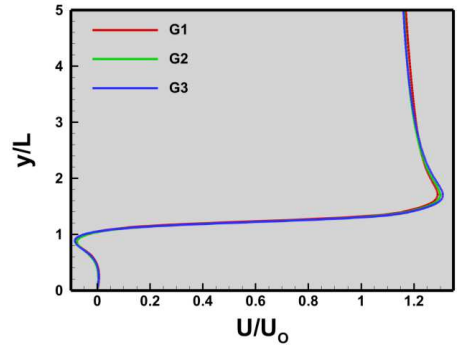

**Figure 5: Effect of grid size on the streamwise velocity profiles.**

Results from Figure 5 indicate a strong grid independence, with all three grids predicting consistent trends for the flow field, with negligible differences between them. Even at the location of maximum discrepancy (y/L=1.7), the normalized velocity differs by <1%. In the wake region of EVG (y/L<1), a noticeable reduction in flow velocity is observed, while above the EVG (y/L>1), the velocity increases and gradually approaches unity as y increases.

To further support the adequacy of the mesh in the entire flow field, we compared non-dimensional instantaneous spanwise vorticity contours ($\omega L/U$) at a representative time step for all three mesh levels (G1, G2, and G3). As shown in Figure 6, all three grids resolve the vortical structures with minimal differences; however, G1 captures the boundaries of these vortices with greater sharpness, offering improved resolution

Additionally, to evaluate how variations in grid resolution influences the fluid-structure interaction (FSI) behavior and the resulting deformation of the tandem EVGs, the sensitivity of the time history of θ to the grid size is analyzed separately for the upstream and downstream EVGs, as shown by Figures 7a and 7b respectively.

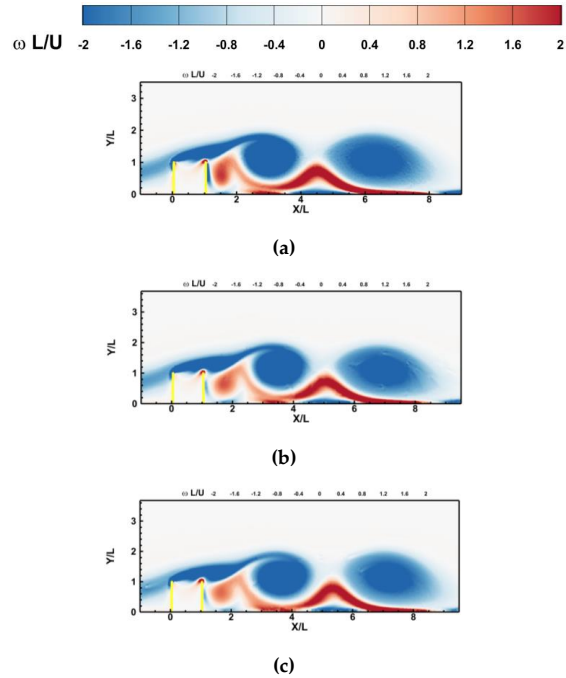

**Figure 6:** Effect of grid size on dimensionless vorticity at a representative time step (a) (G1), (b) Grid 2 (G2), and (c) Grid 3 (G3).

Figure 7 reveals that all the grids predict the general trends for the time histories of θ. A quantitative comparison of the time-averaged inclination angle (θ) shows for the upstream EVG, the average θ values are 89.4878 (G1), 89.5105 (G2), and 89.5419 (G3) degrees. For the downstream EVG, the corresponding values are 90.0498 (G1), 90.0968 (G2), and 90.0722 (G3) degrees. These minimal discrepancies confirm all grids show similar results. However, the G1 grid was chosen for all simulations, ensuring highest accuracy in predicting both the tandem EVG response and the flow field.

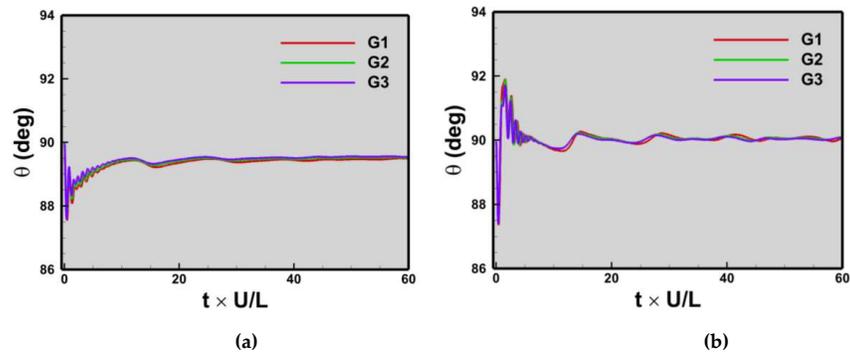

**Figure 7:** Effect of grid resolution on the dynamic responses of tandem EVGs: (a) upstream; and (b) downstream.

Moreover, given the large number of combinations of dimensionless parameters explored in this study, spanning bending rigidity ($0.001 \leq \gamma \leq 5$), mass ratio ($1 \leq \beta \leq 4$), and porosity ($\phi = 0, 0.1, 0.25$, and $0.5$), we adopted a cautious approach by using the finest mesh (G1) for all simulations. This wide parametric range encompasses a broad spectrum of structural flexibility, inertia, and flow permeability, each influencing the FSI behavior in distinct ways. To ensure reliable resolution of the diverse flow-structure interactions across this entire space and eliminate any mesh-induced artifacts, the G1 mesh was used throughout the study.

### 6.3. Time-Step Verification

To evaluate the impact of temporal resolution on the fidelity of the results, we performed a detailed time-step sensitivity study at $\gamma = 0.02$ and $\beta = 1.0$ for $\phi = 0$. Five non-dimensional time steps were tested: $\Delta t_1 = 0.02$ and $\Delta t_2 = 0.025$, $\Delta t_3 = 0.03$, $\Delta t_4 = 0.04$ and $\Delta t_5 = 0.06$. Figures 8a and 8b present the time histories of inclination angle (θ) for the upstream and downstream EVGs, respectively. As shown in Figures 8a and 8b, all time steps reproduce the fundamental dynamic behavior for the upstream and downstream non-perforated EVGs. However, the effect of the time step is pronounced on the amplitude and becomes negligible only when the time step is less than 0.03. Since the predicted oscillation period is approximately $t \times U/L \cong 3.3$ for all simulations as shown in Figure 8, this suggests that there are about 165, 132, 110, 82, and 55 time steps per oscillation period for $\Delta t_1$ through $\Delta t_5$, respectively. Therefore, the time step analysis reveals that the predicted amplitude is sensitive to the time step when fewer than 100 time steps per oscillation period are used. To ensure robust temporal resolution across the entire parametric space of $\gamma$, $\beta$, and $\phi$, we adopted the finest time step, $\Delta t_1 = 0.02$, for all simulations in this study, as the minimum non-dimensional period is about 2.2 for all conditions considered (see Section 6.8).

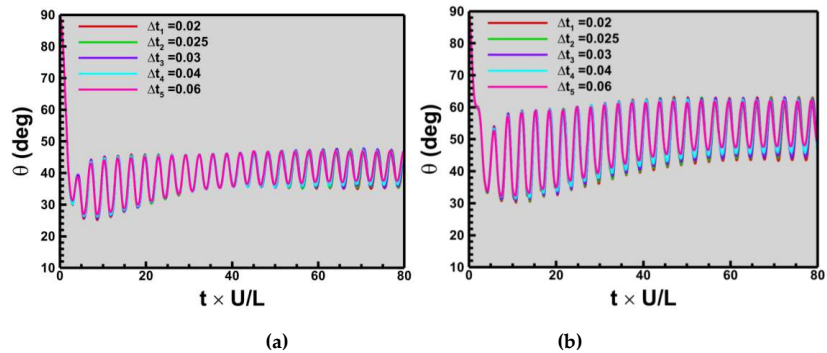

(a)                                                                  (b)

Figure 8: Time-step sensitivity analysis of tandem EVGs: (a) upstream; and (b) downstream.

### 6.4. Non-perforated tandem EVG simulations

To compare our results with those from the perforated tandem EVG simulations, the impact of two dimensionless parameters—dimensionless rigidity and mass ratio—at a fixed Reynolds number of Re=400 was first analyzed on the behavior of non-perforated EVGs. Each parameter is varied independently while the other remains constant, allowing for a clear assessment of their individual effects. Additionally, to have a comprehensive comparison, non-perforated tandem EVG results are compared with non-perforated single EVG results from our previous study[45].

The effect of γ on the dynamic behavior of non-perforated tandem EVG is investigated by fixing β =1.0 and Re = 400. Figure 9a and 9b shows the variation of $\theta_{ave}$ and the $\theta_H$ respectively as a function of γ.

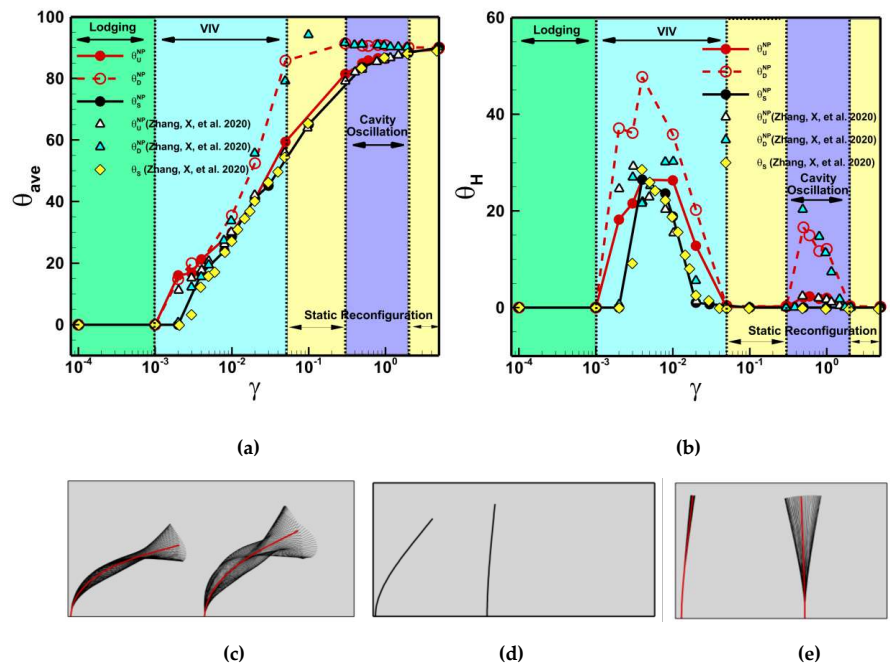

**(a)**                                                                 **(b)**

**(c)**                    **(d)**                    **(e)**

**Figure 9: Dynamic behavior of non-perforated tandem EVGs at different bending rigidities for β=1.0 and Re= 400: (a) $\theta_{ave}$; (b) $\theta_H$; (c) EVG dynamics in VIV mode; (d) EVG dynamics in Static Reconfiguration mode; and (e) EVG dynamics in Cavity Oscillation mode.**

From Figure 9, four different modes of dynamic behavior of the non-perforated tandem EVGs, namely Lodging, VIV, Static Reconfiguration and Cavity Oscillation, are observed. The black solid line represents the non-perforated single EVG case, serving as a reference, while the red solid and dashed lines correspond to the upstream and downstream non-perforated EVGs respectively in a tandem configuration. At low bending rigidity γ ≤ 0.001, the lodging mode is observed (indicated by green color regime), identical to what is observed for single non-perforated EVG, where both the upstream and downstream non-perforated EVGs bends and stays on the ground resulting in zero $\theta_{ave}$ and $\theta_H$. As $\gamma$ increases, at intermediate bending rigidity $0.001 < \gamma < 0.05$, the VIV (vortex-induced vibrations) mode emerges (indicated by blue color regime). This is because the vortices shed periodically from the tip of the non-perforated tandem EVGs, and the shedding frequency at Re = 400 gets close to the second natural frequency of the EVG in that region of $\gamma$, resulting in the lock-in phenomenon, as discussed later in Section 6.8. With increase of bending rigidity $0.05 \leq \gamma \leq 0.3$, $\theta_H$ disappears for both upstream and downstream EVGs and the static reconfiguration mode is observed (indicated by yellow color regime), where both the upstream and downstream EVG bends slightly and stays in that position, identical to what is observed for single non-perforated EVG. Interestingly, as $\gamma$ increase further $0.3 < \gamma < 2$, a unique distinct mode emerges, which is termed as Cavity Oscillation mode (indicated by purple color regime), where the downstream non-perforated EVG bends towards the upstream non-perforated EVG. It is worth noting that this mode does not exist in the case of non-perforated single EVG. This unique mode will be further addressed in detail Section 6.6. At higher bending rigidity $\gamma \geq 2$, $\theta_H$ disappears again and the static reconfiguration mode remerges, with very small bending (large $\theta_{ave}$), identical to what is observed for a single non-perforated EVG. The corresponding behaviors associated with the VIV, Static Reconfiguration, and Cavity Oscillation modes are visualized in Figures 9c–e, respectively.

Overall, $\theta_{ave}$ trend for upstream non-perforated EVG is close to $\theta_{ave}$ trend for the non-perforated single EVG case within the entire range of $\gamma$ investigated, hence suggests that the downstream EVG has a minor effect on the reconfiguration of the upstream EVG. Notably, the downstream EVG exhibits consistently higher $\theta_{ave}$ than the upstream EVG as well as non-perforated single EVG, suggesting that the downstream EVG experiences less bending due to shielding effect from the upstream EVG. Additionally, $\theta_H$ for downstream EVG is always larger than that of the upstream EVG as well as non-perforated single EVG. This is due to the periodic excitation generated by vortex shedding over the upstream EVG, which will be explained later in detail in Section 6.7.

To validate the results, the predicted $\theta_{ave}$ and $\theta_H$ are also compared with published data from Zhang et.al[82] for both single and tandem non-perforated EVGs, as shown in Figure 9. For the single EVG case, excellent agreement is observed in both $\theta_{ave}$ and $\theta_H$, with results closely matching those of Zhang et al.[82] (shown by yellow color diamonds). For the tandem configurations, similarly strong agreement is found for both $\theta_{ave}$ and $\theta_H$, with both studies reporting consistent response modes for upstream and downstream filaments across identical bending rigidity values. However, for the downstream filament, some discrepancies in $\theta_H$ emerge near the peak VIV regime, which needs further investigation.

To analyze how mass ratio might affect the modes predicted in Figure 9, simulations for non-perforated EVGs were repeated for similar cases but with lower mass ratio $\beta = 0.1$, as shown in Figure 10.

From Figure 10, four different modes of dynamic behavior of the non-perforated tandem EVGs, namely Lodging, VIV, Static Reconfiguration and Cavity oscillation are still observed, identical to what was observed for non-perforated tandem EVGs at $\beta = 1.0$ (shown by Figure 9), however the modes are shifted to different $\gamma$ values. This shift is apparent, as the VIV regime for $\beta = 0.1$ has a narrower range, spanning only from $0.001 < \gamma < 0.01$, compared to $0.001 < \gamma < 0.05$ for $\beta = 1$. While the onset remains the same, the transition out of VIV occurs at a lower $\gamma$. Additionally, static reconfiguration appears earlier for $\beta = 0.1$ beginning at $\gamma = 0.01$ compared to $\gamma = 0.05$ for $\beta = 1$ but persists over a narrower range ($0.01 \leq \gamma \leq 0.05$ vs $0.05 \leq \gamma \leq 0.3$) respectively. Moreover, the Cavity Oscillation regime is significantly broadened, starting at $\gamma > 0.05$ for $\beta = 0.1$ compared to γ>0.3 for β =1 and extending to γ < 2 for both cases. Finally, the static reconfiguration mode remerges at γ ≥ 2 for both cases (β =0.1 and β =1). These shifts in modes are directly linked to changes in the EVG's natural frequency, which is inversely related to β. A more detailed discussion on the influence of mass ratio β on the natural frequency and its impact on the transition of dynamic modes is provided in Section 6.8.

In terms of amplitude of oscillation trends, θ_H is significantly reduced in the VIV regime for β =0.1 compared to β =1, whereas in the Cavity Oscillation regime, the amplitude of θ_H is comparable to β =1, due to change of EVG's natural frequency (discussed later in Section 6.8). In terms of θ_ave trends, the lighter EVG (β=0.1) exhibits less bending overall, represented by the higher values of θ_ave, indicating that it remains closer to its original position.

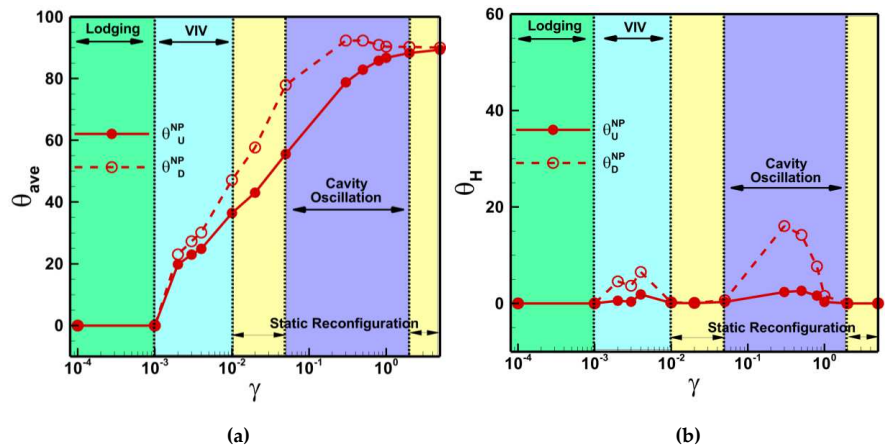

**Figure 10: Dynamic behavior of non-perforated tandem EVGs at different bending rigidities for β=0.1 and Re= 400: (a) θ_ave; and (b) θ_H.**

Overall, the downstream EVG exhibits consistently higher θ_ave and θ_H than the upstream EVG for the entire range of γ investigated at β=0.1, suggesting that the downstream EVG experiences less bending and larger amplitude of oscillation due to the shielding effect and the periodic excitation generated by vortex shedding from the upstream EVG respectively.

The effect of β on the dynamic behavior of the non-perforated tandem EVGs is investigated by fixing γ = 0.5 and Re = 400, as shown by Figure 11. It is worth noting that as per the definition of β, the larger the β, the larger the density of solid, meaning that the structure would be more heavy and thus difficult to retract back to its upright position.

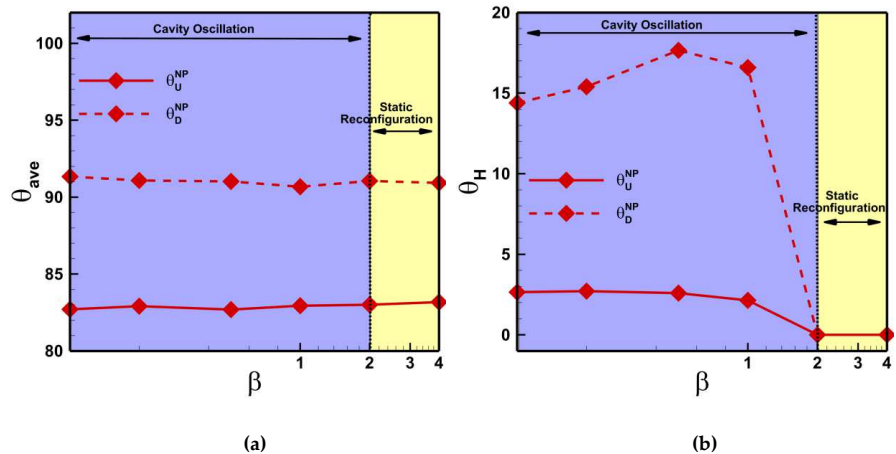

**Figure 11: Dynamic behavior of non-perforated tandem EVGs at different mass ratios for γ = 0.5 and Re= 400: (a) θ$_{ave}$; and (b) θ$_{H}$.**

Unlike previous cases, only two modes are observed as shown in Figure 11. The Cavity Oscillation mode is observed for β < 2, where θ$_{ave}$ remains nearly constant for both upstream and downstream EVGs, with the upstream EVG exhibiting lower values than the downstream EVG. This suggests that the downstream EVG experiences less bending due to shielding effects from the upstream EVG. Meanwhile, θ$_{H}$ shows a clear dependence on β, with the downstream EVG experiences significantly stronger oscillations, reaching its peak at β=0.5 and then decreases with increasing β. As β increases beyond β ≥ 2, both the upstream and downstream EVG start exhibiting Static Reconfiguration mode, where θ$_{H}$ drops to zero, and θ$_{ave}$ remaining constant, indicating that both the upstream and downstream non-perforated EVG bends slightly and stays in that position with zero oscillation. This behavior is attributed to the high structural inertia at large β values, which suppresses dynamic response to unsteady fluid forces. As a result, the EVGs bend slightly under mean fluid loading and remain in a stable, non-oscillatory configuration.

To analyze how bending rigidity affect the modes predicted in Figure 11, simulations were repeated for similar cases but with lower bending rigidity γ = 0.01, as shown in Figure 12.

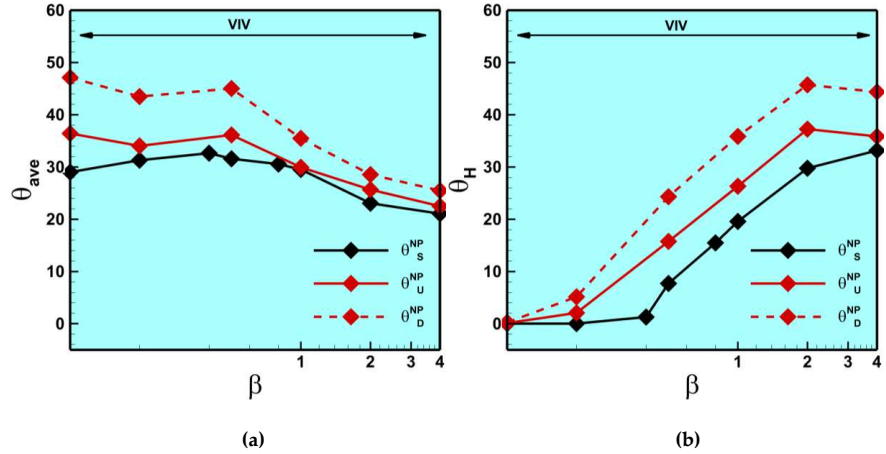

**Figure 12: Dynamic behavior of non-perforated tandem EVGs at different mass ratios for $\gamma = 0.01$ and Re= 400: (a) $\theta_{ave}$; and (b) $\theta_H$.**

From Figure 12, the static configuration mode is observed for $\beta \le 0.1$, where the non-perforated tandem EVG bends slightly and stays in that position. As $\beta > 0.1$, both the upstream and downstream non-perforated EVGs starts exhibiting the VIV mode, as the vortex shedding frequency gets close to the second natural frequency in that range of $\beta$, as discussed later in Section 6.8. Overall, the downstream EVG consistently shows higher $\theta_{ave}$ than the upstream EVG as well as non-perforated single EVG, indicating that the upstream structure bends more due to direct exposure to the flow, while the downstream EVG experiences less bending due to shielding effects. However, in terms of oscillations, the downstream EVG undergoes higher $\theta_H$ across all $\beta$ investigated, surpassing both the upstream and single EVG due to wake-induced amplification effects from the upstream EVG. Compared to trends observed at fixed $\gamma$=0.5, where Cavity Oscillation was observed for $\beta < 2$ (Figure 11), interestingly no Cavity Oscillation mode is observed at fixed $\gamma$=0.01, and the system remains locked in VIV mode across all $\beta$. $\theta_{ave}$ and $\theta_H$ are expected to eventually reach nearly zero at very large $\beta$ values demonstrating the lodging configuration due to the heaviness of the EVGs.

### 6.5. Perforated tandem EVG simulations

We performed simulations across various perforation levels while keeping bending rigidity constant to isolate the effects of porosity. Since perforation changes EI, keeping bending rigidity constant requires adjusting EI values. In the literature [96,113], an equation for equivalent bending stiffness (EI$_P$) for perforated filament has been developed theoretically as follows:

$$EI_P = KEI_{NP} \qquad (6)$$

where

$$K = \frac{(N+1)B(N^2+2N+B^2)}{(1-B^2+B^3)N^3+3BN^2+(3+2B-3B^2+B^3)B^2N+B^3} \qquad (7)$$

Here, $B = \frac{b}{b+a}$; N is number of holes, $EI_{NP}$ is bending stiffness for non-perforated structures, $EI_P$ is bending stiffness for perforated structures. Equation 7 suggests that the bending stiffness for the perforated filaments reduces by a factor K due to the presence of pores. Therefore, we increased the flexural modulus of perforated cases by the factor $1/K$ to have a similar rigidity for perforated and non-perforated filaments.

The perforated tandem EVG simulations are conducted for various $\phi$= [0.1,0.25, and 0.5]. The effect of $\gamma$ on the dynamic behavior of perforated tandem EVGs is investigated by fixing $\beta$ =1.0 and Re = 400, as shown by Figure 13. From Figure 13, the most significant finding for different porosities $\phi$= [0.1,0.25, and 0.5] is the complete absence of Cavity Oscillation mode in case of perforated tandem EVGs compared to the non-perforated tandem EVGs (shown by Figure 9) at $\beta$ =1.0. For the non-perforated tandem EVG configuration, a well-defined Cavity Oscillation regime appeared for $0.3 < \gamma < 2$, whereas in the perforated tandem EVG configuration, this mode is entirely suppressed for all porosities investigated. Instead, the system transitions directly from VIV to Static Reconfiguration mode without undergoing secondary oscillations. Additionally, increasing perforation leads to a shift in different modes. This shift is evident from the narrowing of the VIV regime and the broadening of the static reconfiguration regime as perforation increases. This shift is attributed to modifications in natural frequency of the perforated structures, which alters the resonance conditions, as discussed in detail later in Section 6.8.

Furthermore, an increase in perforation leads to notable changes in both $\theta_{ave}$ and $\theta_H$ across all regimes. In the static reconfiguration mode, $\theta_{ave}$ for perforated tandem EVGs remains close to $\theta_{ave}$ of non-perforated tandem EVGs, suggesting that bending in the static reconfiguration regime is largely unaffected by change in porosity. However, in the VIV regime, $\theta_{ave}$ for perforated tandem EVGs increases with increase in porosity, indicating that increase in perforation reduces bending of both the upstream and downstream EVGs. Additionally, the amplitude of oscillation $\theta_H$ consistently decreases with increasing porosity, further confirming that perforations shifts/modifies the natural frequency of the structure, moving it away from resonance with vortex shedding and reducing the oscillatory energy by damping the motion. Since non-perforated and perforated tandem EVGs with similar mechanical properties exhibit different responses, it can be concluded that perforation affects the fluid loads on the tandem EVGs, as explored in the next Section 6.5.1.

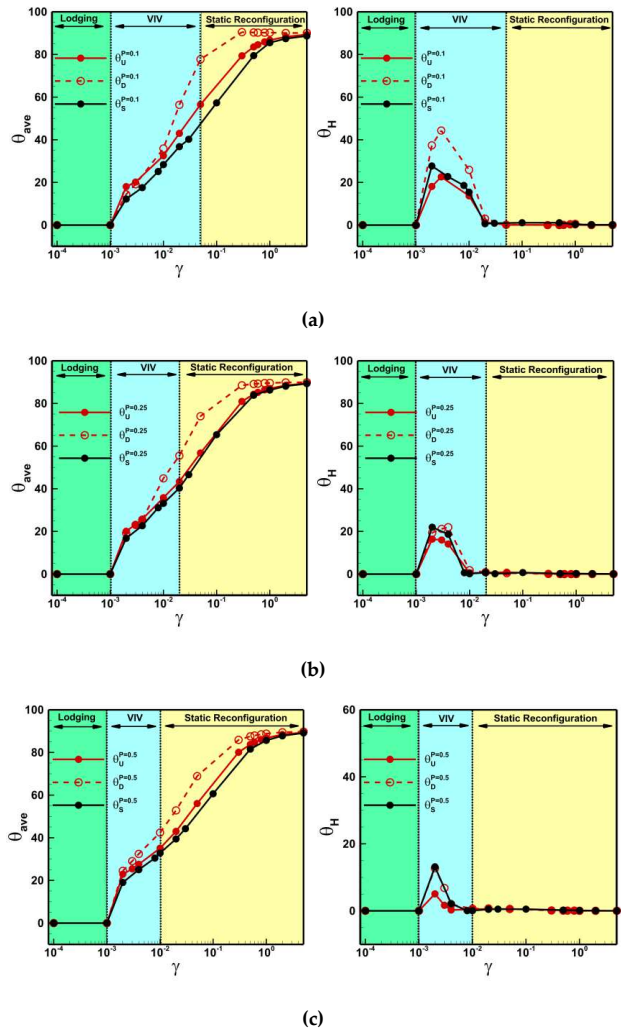

**Figure 13: Dynamic behavior of perforated tandem EVGs at different bending rigidities for $\beta$=1.0 and Re= 400: (a) $\phi$ = 0.1; (b) $\phi$ = 0.25; and (c) $\phi$ = 0.5.**

The effect of γ on the dynamic behavior of perforated tandem EVGs at β =0.1 ais shown in Figure 14.

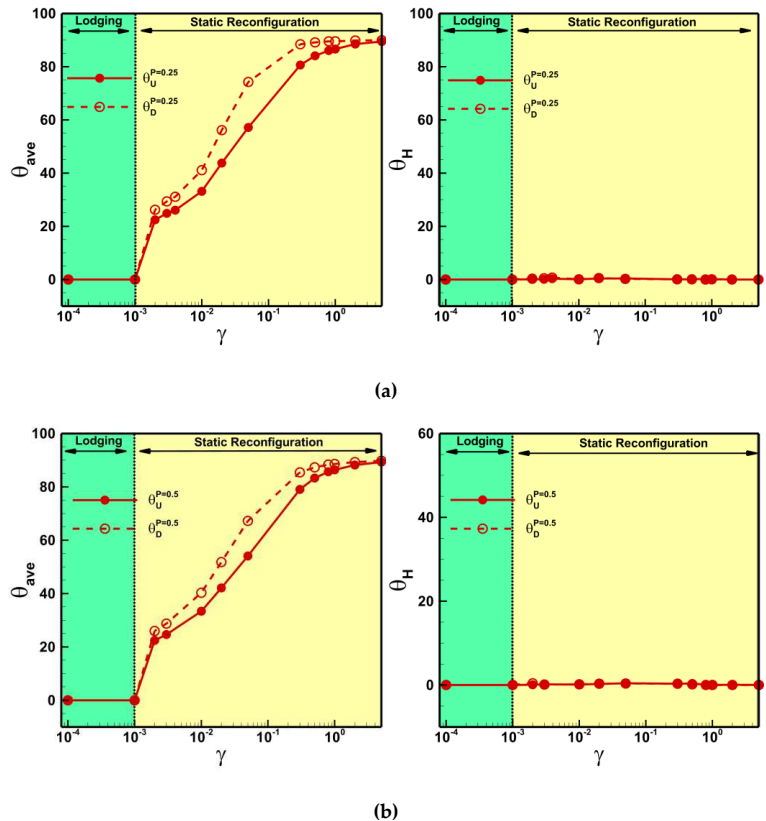

**(a)**

**(b)**

**Figure 14: Dynamic behavior of perforated tandem EVGs at different bending rigidities for *β*=0.1 and Re= 400: (a) ϕ = 0.25; (b) ϕ = 0.5.**

From Figure 14, it is apparent that increasing perforation fundamentally alters the dynamic response of both the upstream and downstream EVGs by eliminating different oscillatory modes observed for non-perforated tandem EVGs at β = 0.1 (shown by Figure 10). Unlike the non-perforated tandem EVG case, where VIV and Cavity Oscillation were observed, perforated cases at ϕ= [0.25, and 0.5] transition directly from Lodging to Static Reconfiguration mode without any oscillations. The complete suppression of $\theta_H$ suggests that perforation not only modifies the EVG's natural frequency but also introduces damping that inhibit the oscillatory motion of the EVG, as discussed in detail in Section 6.8.

Furthermore, it can be observed from Figure 14 that the perforation influences structural deformation differently for the upstream and downstream EVGs. For the upstream EVG, $\theta_{ave}$ slightly increases with increase in perforation, indicating less bending with increase in porosity. Conversely, the downstream EVG shows a decreasing trend for $\theta_{ave}$, suggesting greater bending with increase in porosity. This contrast trend highlights that perforations alter the fluid loading on both the upstream and downstream EVG differently, as explored in the next Section 6.5.1.

The effect of β on the dynamic behavior of the perforated tandem EVGs is investigated by fixing γ = 0.5 and Re = 400, as shown in Figure 15.

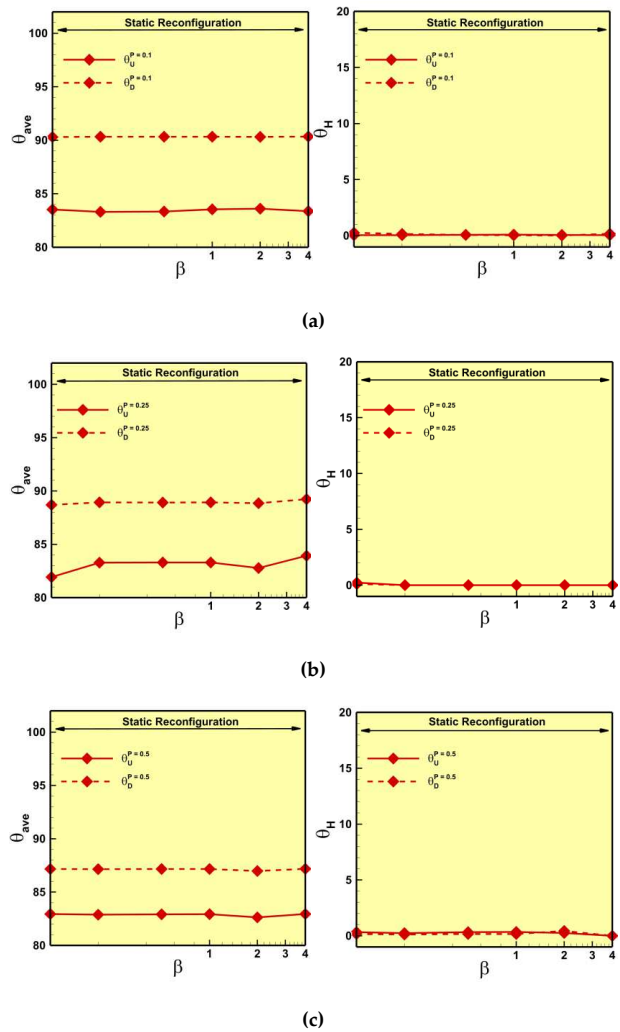

**(a)**

**(b)**

**(c)**

Figure 15: Dynamic behavior of perforated tandem EVGs at different mass ratios for

γ = 0.5 and Re= 400: (a) φ = 0.1; (b) φ = 0.25; and (c) φ = 0.5.

Figure 15 depicts that increasing porosity ($\phi$=0.1,0.25,0.5) eliminates the Cavity Oscillation mode completely, which was observed for non-perforated tandem EVGs (Figure 11), transitioning the system entirely into Static Reconfiguration. For non-perforated tandem EVGs, Cavity Oscillation was observed for $\beta$<2 with significant oscillations, particularly in the downstream EVG. However, for perforated cases, $\theta_H$ remains near zero across all $\beta$ investigated, confirming a complete suppression of unsteady oscillatory motion. This absence is attributed to the change in the natural frequency of the perforated structures, which shifts them away from resonance, and to the added damping introduced by perforations, as discussed in detail later in Section 6.8.

The structural deformation trends also shift with increase in perforation, with $\theta_{ave}$ sometimes increasing and sometimes decreasing for the upstream EVG, whereas $\theta_{ave}$ decreasing for the downstream EVG with increase in perforation, suggesting greater bending. This indicates that perforations induce complex modifications in altering the fluid loading on both the upstream and downstream EVGs differently, as explored in the next Section 6.5.1.

The effect of $\beta$ on the dynamic behavior of the perforated tandem EVGs at $\gamma = 0.01$ ais shown in Figure 16.

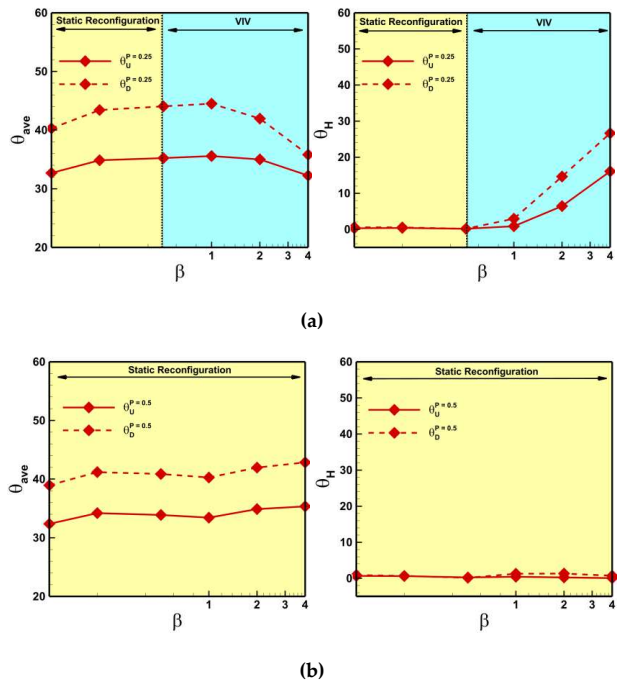

**Figure 16: Dynamic behavior of perforated tandem EVGs at different mass ratios for $\gamma = 0.01$ and Re= 400:(a) $\phi$ = 0.25; and (b) $\phi$ = 0.5.**

From Figure 16, it is observed that for fixed $\gamma$=0.01 and Re=400, introducing perforations leads to a progressive shift in dynamic behavior of both upstream and downstream EVG in tandem configuration. In the non-perforated tandem EVG case (shown by Figure 12), both Static Reconfiguration and VIV modes are observed, but as porosity increases to $\phi$=0.25, VIV mode is significantly narrowed, and at $\phi$=0.5, it is entirely eliminated, leaving only the Static Reconfiguration mode across all $\beta$.

The trends in $\theta_{ave}$ for the upstream and downstream EVGs do not exhibit a monotonic dependence on perforation, with values sometimes increasing and sometimes decreasing compared to the non-perforated case. Additionally, $\theta_{H}$ systematically decreases with increasing porosity for both upstream and downstream tandem perforated EVG, confirming the suppression of oscillatory behavior or VIV mode.

Figure 17 depicts the map summarizing the distribution of different dynamic response regimes observed in the two-dimensional ($\gamma$, $\beta$) parameter space for tandem EVGs with varying porosity of $\phi$ = 0, 0.1, 0.25, and 0.5 respectively. For non-perforated EVGs ($\phi$ = 0), all dynamic behavior modes including Lodging, Vortex-Induced Vibration (VIV), Cavity Oscillation, and Static Reconfiguration are observed at both fixed $\beta$ = 1 and $\beta$ = 0.1. However, the location and extent of these regimes shift between the two cases due to changes in natural frequency of the EVGs as structural inertia (solid density) changes, which will be discussed in detail in Section 6.8. With increasing porosity, Cavity Oscillation mode is entirely eliminated for all porosities, confirming that cavity Oscillation mode is distinct from VIV mode, as VIV still persists at all porosity levels for $\beta = 1$, even though Cavity Oscillation mode is not. Additionally, a progressive shift in dynamic regimes is observed with increasing porosity, again attributing to modifications in the natural frequency of the perforated EVGs compared to non-perforated EVGs. Along fixed $\gamma = 0.01$, the non-perforated case exhibits both Static and VIV responses; however, as porosity increases, the VIV region gradually narrows and is fully suppressed at higher porosity of $\phi = 0.5$, where only static reconfiguration is observed. For fixed $\gamma = 0.5$, Cavity Oscillation and Static Reconfiguration mode are observed at $\phi = 0$, but with increasing porosity, the Cavity Oscillation mode disappears entirely, leaving only static reconfiguration mode for all porosities. These results collectively demonstrate that porosity not only eliminates cavity oscillation mode but also drives mode shifting across the parameter space, primarily through its effect on the structure's (EVG's) natural frequency and additional damping introduced by perforations. This will be further discussed in detail in Section 6.8.

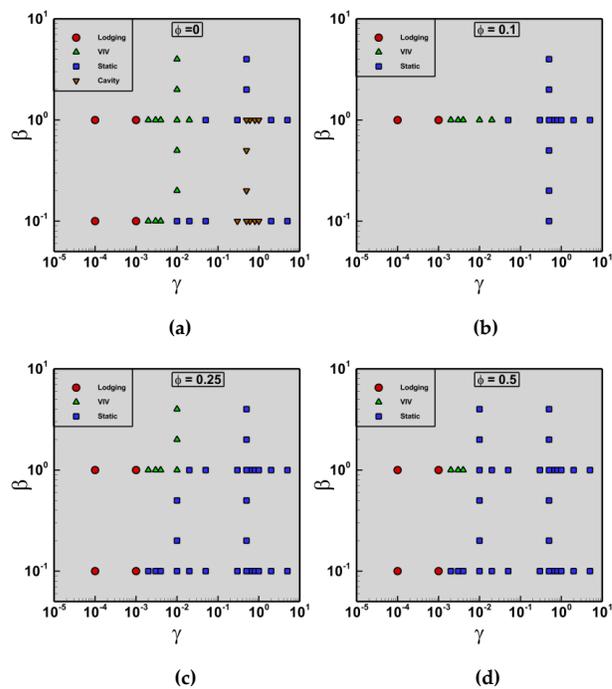

**Figure 17: Maps of dynamic response regimes in the two-dimensional ($\gamma$, $\beta$) parameter space for a tandem configuration of EVGs for fixed Re=400 and D=1L, shown for varying porosity levels: (a) $\phi$ = 0; (b) $\phi$ = 0.1; (c) $\phi$ = 0.25; and (d) $\phi$ = 0.5.**

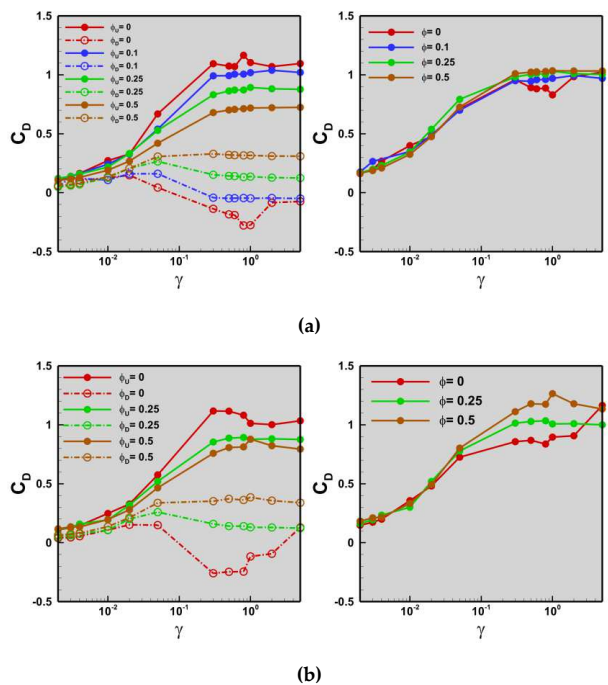

**(a)**

**(b)**

**Figure 18. Drag coefficients of tandem EVGs at different porosity levels: (a) Drag coefficient of each EVG (left) and Total drag (right) for $\beta = 1$; and (b) Drag coefficient of each EVG (left) and Total drag (right) for $\beta = 0.1$**

### 6.5.1. Effect of perforation on fluid loads

Following the analysis of regime transitions in the ($\gamma$, $\beta$) parameter space, we now investigate the impact of porosity on fluid loading, with a focus on the evolution of drag forces. Drag data for the configurations presented in Figure 17 are examined and the corresponding drag coefficient ($C_D = \frac{F_D}{\frac{1}{2}\rho_f U_\infty^2 A}$) is reported in Figures 18 and 19. Here $F_D$ denotes the mean drag force, $\rho_f$ is the fluid density, $U_\infty$ is the inlet velocity, and A=L×w represents the frontal area of the EVG.

From Figure 18a and Figure 18b, it is apparent that for non-perforated tandem EVGs (shown by red color line), the upstream EVG consistently experiences higher drag than the downstream EVG across both $\beta = 1$ and $\beta = 0.1$ cases respectively. This is attributed to the wake shielding effect, where the upstream EVG blocks the incoming flow and generates a momentum-deficient wake, significantly reducing the fluid loading on the downstream EVG compared to the upstream EVG. As porosity increases, the drag on the upstream EVG decreases due to bleeding flow through pores, which lowers the pressure differential across its surface and thus weakens the wake. In contrast, drag on the downstream EVG increases with increasing porosity for most of the range of $\gamma$ investigated, as more flow is transmitted through the upstream EVG, allowing the downstream EVG to interact with more flow.

It is worthwhile to note that for non-perforated tandem EVGs, the downstream EVG experiences negative drag under the presence of cavity oscillation mode. This occurs due to the creation of a low-pressure region caused by the strong, recirculating wake formed by the upstream EVG, which pulls the downstream EVG towards upstream EVG, hence exhibiting negative drag. Introducing porosity eliminates this cavity oscillation behavior completely by allowing fluid transmission through the upstream EVG via bleeding flow, thus restoring streamwise momentum and resulting in stable, positive drag on the downstream EVG.

Moreover, Figures 18a and Figure 18b indicate that total drag increases with increase in bending rigidity ($\gamma$), as both upstream and downstream EVGs become more rigid and less responsive to flow-induced deformation. Although porosity appears to increase total drag under certain conditions, these instances correspond to shifts in dynamic regimes or modes, where perforated and non-perforated EVGs no longer exhibit comparable dynamic motion behavior, thereby complicating direct comparisons. For instance, at $\gamma$ =1 for fixed $\beta = 1$ (Figure 18a), the non-perforated tandem EVG exhibits Cavity Oscillation, whereas introducing porosity ($\phi$=0.1,0.25,0.5) causes a complete transition to Static Reconfiguration (refer to Figure 17). In such cases, the dynamic behavior changes entirely, making a direct evaluation of drag misleading. However, when the dynamic mode remains the same—for example, in the VIV regime at $\gamma$=0.01, increasing porosity consistently leads to a reduction in total drag by damping large-amplitude oscillations and weakening vortex-induced forces. This distinction underscores the importance of comparing drag trends within the same dynamic regime.

Figure 19a and 19b depicts the drag trends for $\gamma = 0.5$ and $\gamma = 0.01$ respectively, and the results are consistent with the observations from Figure 18. It is apparent that the upstream EVG always experiences higher drag than the downstream EVG, and increasing porosity reduces upstream drag while increasing downstream drag, due to the combined effects of wake weakening and flow transmission through the upstream EVG due to bleeding flow. However, a key distinction arises in the presence of cavity oscillation mode. At $\gamma = 0.5$, where the structure is more rigid, the non-perforated downstream EVG exhibits negative drag, indicating the formation of a low-pressure recirculating wake and the onset of cavity oscillation behavior. In contrast, at $\gamma = 0.01$, where the structure is comparably flexible, the cavity oscillation

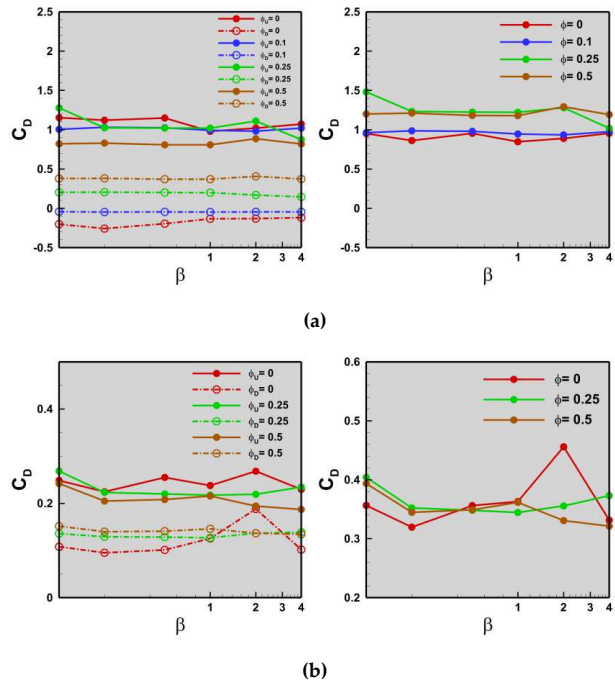

**(a)**

**(b)**

**Figure 19:** Drag coefficients of tandem EVGs at different porosity levels: (a) Drag coefficient of each EVG (left) and Total drag (right) for $\gamma = 0.5$; and (b) Drag coefficient of each EVG (left) and Total drag (right) for $\gamma = 0.01$.

mode is absent, and the drag on the downstream EVG remains positive even in the non-perforated EVG case. This difference highlights the role of bending rigidity in enabling or suppressing suction-driven dynamics: a stiffer structure supports the formation of a confined wake conducive to cavity oscillations, whereas increased flexibility leads to large deformations that disrupt the coherent pressure field required for cavity formation. In terms of total drag shown by Figure 19a and 19b, $\gamma = 0.5$ results in a higher overall drag across all porosities due to greater rigidity, while $\gamma = 0.01$ exhibits lower total drag as the flexible EVGs align more closely with the flow, hence minimizing resistance.

### 6.6. Dynamic behavior of tandem EVGs

To gain deeper insight into the oscillatory dynamic behavior of tandem EVGs, we analyze their dynamics within two distinct regimes: the Vortex-Induced Vibration (VIV) regime and the Cavity Oscillation mode. These regimes are characterized using the time history, phase portrait, and Fast Fourier Transform (FFT) of θ. Firstly, the focus is laid on how γ influences dynamic motion, by fixing β = 1.0 and Re = 400. Two representative cases are selected: γ = 0.5 (shown by Figure 20 and Figure 21), which corresponds to cavity oscillations, and γ = 0.01 (shown by Figure 22), which exhibits VIV behavior respectively. These cases were chosen because they represent distinct dynamic modes and also lie at the intersection of fixed-parameter lines in the dynamic regime map (shown by Figure 17), making them ideal for comparative analysis. Both non-perforated (φ = 0) and perforated EVGs (φ = 0.25) are examined to highlight the role of porosity in modifying flow-induced structural oscillations.

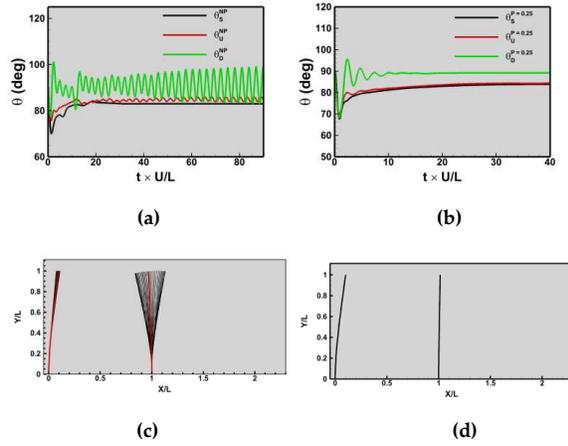

**Figure 20: Dynamic behavior of tandem EVGs for varying porosity levels at γ = 0.5 for fixed β = 1.0 and Re = 400: (a) Time history of θ for φ =0; (b) Time history of θ for φ =0.25; (c) EVG dynamics for φ =0; and (d) EVG dynamics for φ =0.25.**

Figure 20a shows the time histories of bending of the upstream (shown by red color) and downstream EVGs (shown by green color) in the non-perforated tandem configuration at γ = 0.5 for fixed β = 1. It is apparent that the non-perforated tandem EVGs bend initially from its upright position (θ=90 degrees) showing significant initial bending and then a dynamic oscillatory behavior with an amplitude of oscillation is observed with time passing by. The downstream EVG exhibits large-amplitude periodic oscillations, while the upstream EVG shows small fluctuations and closely resembles the response of a single EVG. This behavior is indicative of the cavity oscillation mode, where the downstream EVG is driven by suction forces within a low-pressure, recirculating wake behind the upstream EVG. The oscillation pattern reflects a back-and-forth motion, as the downstream EVG moves alternately toward and away from the upstream EVG, consistent with the unsteady suction–recovery cycle. This behavior can be better visualized in Figure 20c, where the downstream EVG shows a back-and-forth motion, while the upstream EVG shows very small oscillations, with the average deflection profile shown in red. It should be noted that the unsteady suction-recovery cycle is due to the pressure fluctuation in the cavity. The pressure fluctuations arise from the cyclic instability of the free shear layer formed by flow separation at the cavity's upstream edge. This shear layer is inherently unstable and amplifies small disturbances as it convects across the cavity, generating concentrated vortices. When these vortices impinge on the downstream edge of the cavity, they produce strong pressure fluctuations within the cavity. These fluctuations, in turn, perturb the shear layer at the upstream edge, seeding new instabilities and completing a feedback loop. This self-sustaining cycle leads to persistent shear layer instabilities and cavity oscillations.

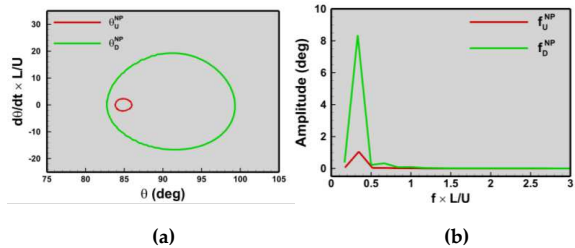

**Figure 21: Dynamic behavior of a non-perforated tandem EVGs at γ = 0.5 for fixed β = 1.0 and Re = 400: (a) Phase portrait; and (b) Fast Fourier Transform (FFT).**

In contrast, for the perforated configuration ($\phi = 0.25$) shown by Figure 20b, both upstream and downstream tandem EVGs exhibit an initial bending but quickly converge to steady orientations (showing static reconfiguration mode) with negligible oscillations. This suppression of motion results from bleeding flow through the upstream EVG, which weakens the cavity by restoring streamwise momentum and reducing suction-driven excitation, discussed in detail in Section 6.7.

Figure 21a further supports this behavior through the phase portrait for $\gamma = 0.5$. The elliptical shape phase diagram indicates a stable dynamic state. The downstream non-perforated EVG exhibits a larger and wider elliptical loop, indicative of larger amplitude of oscillations, while the upstream non-perforated EVG forms relatively smaller loop.

The Fast Fourier Transform (FFT) plot in Figure 21b aligns with these observations. The downstream EVG in the non-perforated case exhibits a strong first harmonic peak at the dimensionless frequency of 0.34, while the upstream EVG shows a relatively smaller peak at same frequency. There also exists, a second peak (second harmonic) for non-perforated tandem EVGs, however they are not as prominent as the first peak. The frequency characteristics of the Cavity Oscillation mode across the full range of $\gamma$ and $\beta$ investigated will be discussed in detail in Section 6.8.

At γ = 0.01, the non-perforated tandem EVGs (Figure 22a) exhibit classic vortex-induced vibration (VIV) behavior, with both the upstream and downstream EVGs undergoing large-amplitude, oscillations. The upstream EVG closely resembles the response of the single EVG, while the downstream EVG exhibits greater oscillation amplitude, enhanced by wake-induced excitation from the upstream EVG. In contrast to the γ = 0.5 case (Figure 20 and Figure 21), where only the downstream EVG oscillates due to cavity suction, the γ = 0.01 configuration produces larger, symmetric oscillations involving both EVGs, reflecting stronger fluid–structure interaction enabled by greater flexibility. This dynamic behavior is further visualized in Figure 22g, where both EVGs display a stationary sinusoidal oscillatory pattern, with the average deflection profile shown in red. The shape of the deformation indicates a second mode of oscillation, as characterized by a sinusoidal wave along the EVG span in both upstream and downstream filaments.

The phase portraits (Figure 22c) reinforce this distinction. Both EVGs trace asymmetric elliptical loops, characteristic of harmonic motion, with the downstream EVG forming a significantly larger loop—a direct indication of its greater angular amplitude. This difference aligns with the time-history results and is further supported by the FFT analysis in Figure 22e. The upstream EVG, which exhibited near-static reconfiguration at γ = 0.5, now displays clear periodic motion at γ = 0.01. This contrast demonstrates that lower bending rigidity activates dynamic response in both EVGs, while higher rigidity confines motion to the downstream EVG.

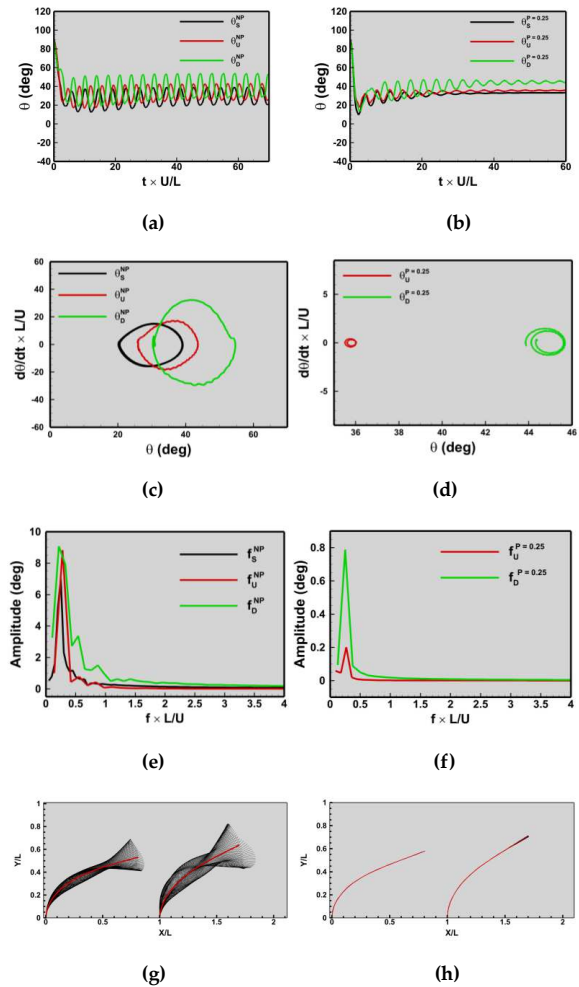

**Figure 22: Dynamic behavior of a tandem EVGs for varying porosity levels at γ = 0.01 for fixed β = 1.0 and Re = 400: (a) Time history of θ for φ =0; (b) Time history of θ for φ =0.25; (c) Phase portrait for φ=0; (d) Phase portrait for φ =0.25; (e) Fast Fourier Transform (FFT) for φ =0; (f) Fourier Transform (FFT) for φ =0.25; (g) EVG dynamics for φ =0; and (h) EVG dynamics for φ =0.25.**

The FFT spectra (Figure 22e) confirm this interpretation. A strong primary frequency peak near fL/U ≈ 0.35 appears for both EVGs, with the downstream EVG showing the highest amplitude, consistent with its larger phase loop. Additionally, the downstream EVG exhibits weaker second and third harmonic peaks, suggesting mild nonlinearities—though these are far less prominent than the dominant mode. In contrast, the γ = 0.5 case (Figure 20 and Figure 21) exhibited a weaker, lower-frequency peak, limited to the downstream EVG, confirming that VIV at low γ results in stronger, system-wide oscillations, unlike the localized, lower-energy cavity oscillation observed at higher γ.

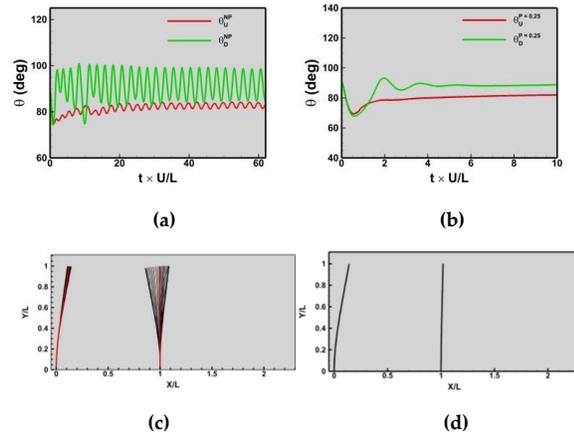

**(a)** **(b)**

**(c)** **(d)**

**Figure 23: Dynamic behavior of a tandem EVGs for varying porosity levels at β = 0.1 for fixed γ =0.5 and Re = 400: (a) Time history of θ for ϕ =0; (b) Time history of θ for ϕ =0.25; (c) EVG dynamics for ϕ =0; and (d) EVG dynamics for ϕ =0.25.**

For perforated tandem EVGs (ϕ = 0.25; Figure 22, the oscillation amplitude is significantly reduced, especially for the upstream EVG, which becomes nearly static and closely resembles the trend for single perforated EVG. The downstream EVG retains weak periodic motion, but with a reduced amplitude of oscillation and frequency, as reflected in its smaller phase loop and diminished FFT peak. The disappearance of secondary harmonics also suggests further suppression of nonlinear dynamics due to porosity. While porosity completely eliminates cavity oscillations at γ = 0.5, here it attenuates, but does not eliminate VIV mode especially for the downstream EVGs.

To understand the effect of β on the dynamic behavior of the non-perforated and perforated tandem EVGs, Figure 23 and 24 shows the time history, phase portrait, and FFT of the inclination angle at β =

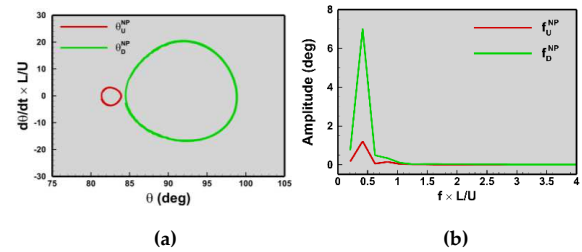

**(a)** **(b)**

**Figure 24: Dynamic behavior of a non-perforated tandem EVGs at β = 0.1 for fixed γ=0.5 and Re = 400: (a) Phase portrait; and (b) Fast Fourier Transform (FFT).**

0.1 for fixed $\gamma$=0.5. The results are also compared with $\gamma$ = 0.5 at fixed $\beta$ = 1 (shown in Figure 20 and Figure 21), to illustrate the influence of mass ratio. For the non-perforated tandem EVG configurations, both cases exhibit the cavity oscillation mode, with the downstream EVG undergoing large-amplitude periodic motion driven by suction forces in the wake of the upstream EVG. However, at $\beta$ = 0.1, the amplitude of oscillation in the downstream EVG is reduced compared to $\beta$ = 1, indicating that increased structural inertia enhances cavity-driven unsteadiness or oscillations. The upstream EVG, however, shows comparable oscillation amplitude across both $\beta$ values. This trend is reflected in the phase portraits, where the downstream EVG loop is smaller at $\beta$ = 0.1, and in the FFT spectra, which show lower primary peak amplitude and less pronounced secondary harmonics, confirming a weaker nonlinear response at lower $\beta$=0.1.

Additionally, it is apparent from Figure 23 and Figure 24, that perforated tandem EVGs ($\phi$ = 0.25) at $\beta$ = 0.1, both upstream and downstream EVGs exhibit static reconfiguration, and oscillations are suppressed entirely. The time history shows a steady initial bending with no sustained unsteady oscillatory motion. As a result, no phase portrait or frequency spectrum is presented for this case. The same trend was observed for perforated tandem EVGs at $\beta$ = 1. Therefore, these results emphasize that cavity oscillation is highly sensitive to porosity.

### 6.7. Local flow study

To gain further insight into the local flow behavior of non-perforated and perforated tandem EVGs, we examine the vortical structures generated around them to understand how different oscillation modes, specifically vortex-induced vibration (VIV) and the cavity oscillation mode, alter the wake dynamics. To capture representative behaviors, three cases are considered: first, for fixed $\beta = 1$, two values of bending rigidity are examined, $\gamma = 0.5$, corresponding to cavity oscillation, and $\gamma = 0.01$, corresponding to VIV. Then, for fixed $\gamma = 0.5$, a lower mass ratio of $\beta = 0.1$ is selected to study the influence of inertia/solid density on the dynamic behavior of non-perforated and perforated tandem EVGs. For each case, both non-perforated ($\phi = 0$) and perforated ($\phi = 0.25$) configurations are analyzed to assess how porosity alters the wake structure and modifies the flow-induced excitation mechanisms.

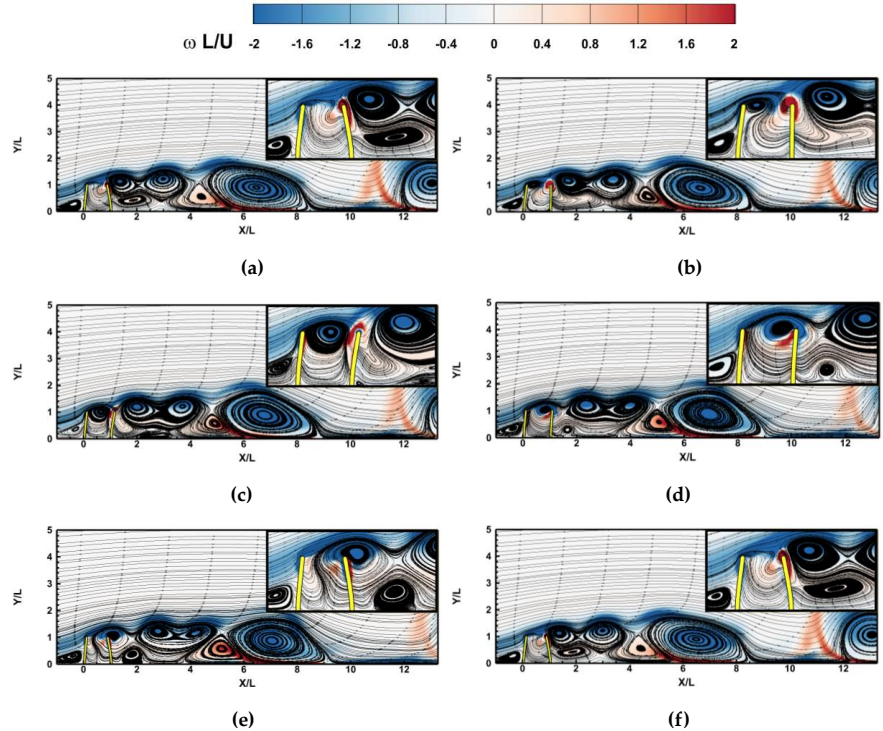

**Figure 25: Dimensionless vorticity and streamlines plots representing wake structures of a non-perforated tandem EVG ($\phi = 0$), at $\gamma = 0.5$ for fixed $\beta = 1$ and Re=400, shown at different instances: (a) t/T=0; (b) t/T=0.2; (c) t/T=0.4; (d) t/T=0.6 (e) t/T=0.8; and (f) t/T= 1.0.**

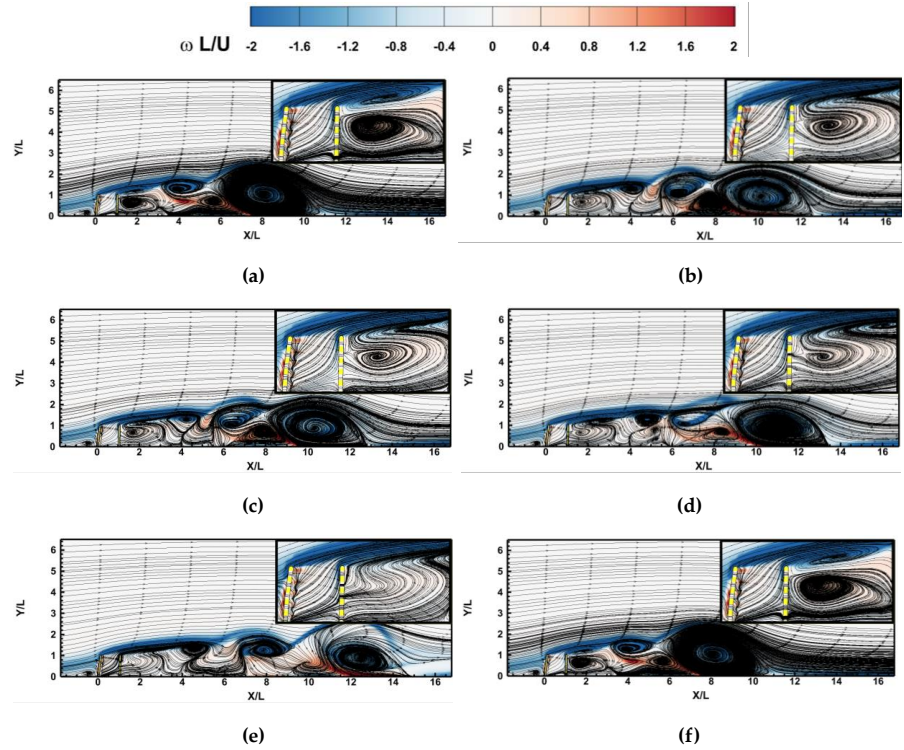

**Figure 26: Dimensionless vorticity and streamlines plots representing wake structures of a perforated tandem EVG ($\phi$ = 0.25), at $\gamma$ = 0.5 for fixed $\beta$ = 1 and Re=400, shown at different instances: (a) t/T=0; (b) t/T=0.2; (c) t/T=0.4; (d) t/T=0.6 (e) t/T=0.8; and (f) t/T= 1.0.**

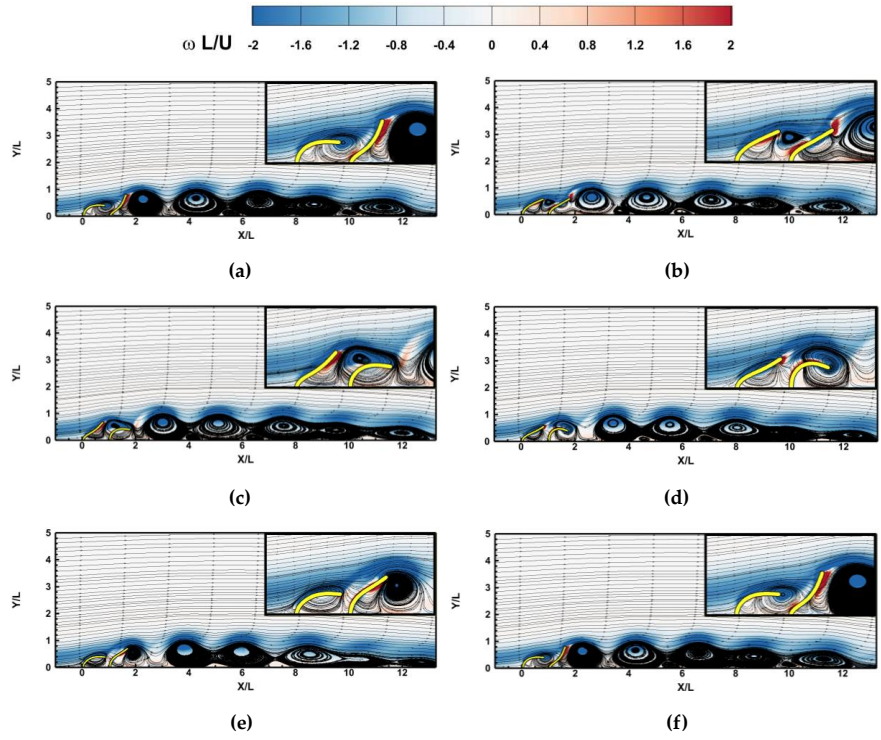

**Figure 27:** Dimensionless vorticity and streamlines plots representing wake structures of a non-perforated tandem EVG ($\phi = 0$), at $\gamma = 0.01$ for fixed $\beta = 1$ and Re=400, shown at different instances: (a) t/T=0; (b) t/T=0.2; (c) t/T=0.4; (d) t/T=0.6 (e) t/T=0.8; and (f) t/T= 1.0.

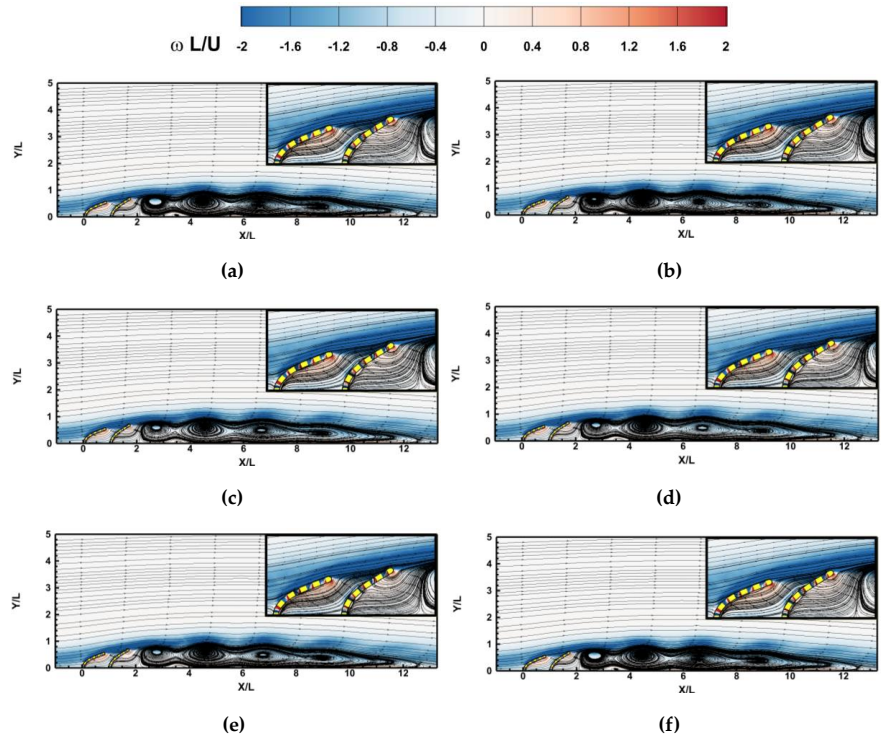

**Figure 28: Dimensionless vorticity and streamlines plots representing wake structures of a perforated tandem EVG (ϕ = 0.25), at γ = 0.01 for fixed β = 1 and Re=400, shown at different instances: (a) t/T=0; (b) t/T=0.2; (c) t/T=0.4; (d) t/T=0.6 (e) t/T=0.8; and (f) t/T= 1.0.**

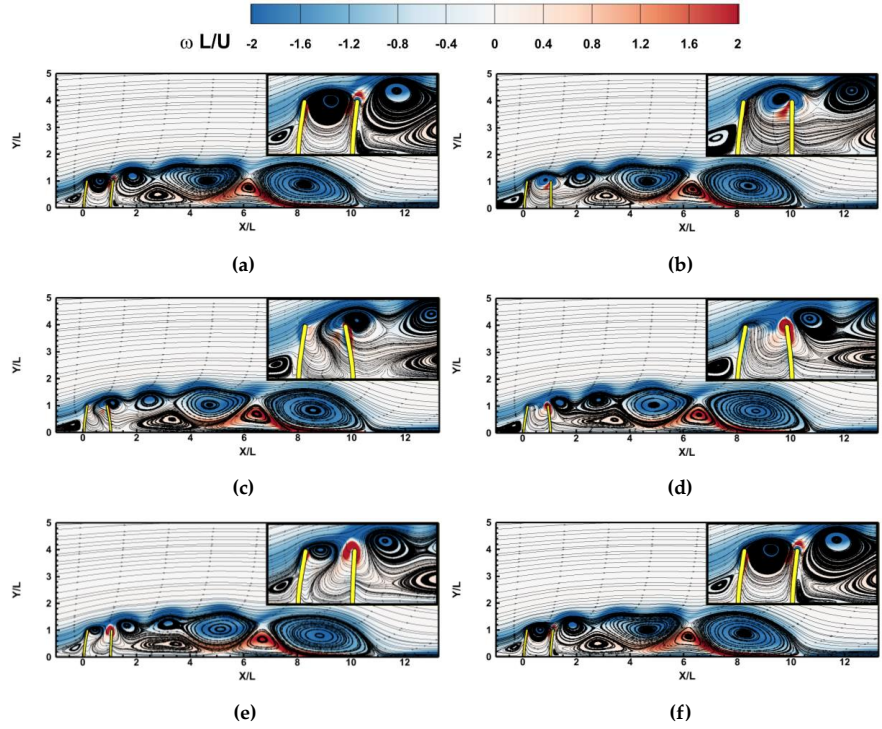

**Figure 29:** Dimensionless vorticity and streamlines plots representing wake structures of a non-perforated tandem EVG ($\phi$ = 0), for $\beta$ = 0.1 at fixed $\gamma$ = 0.5 and Re=400, shown at different instances: **(a)** $t/T$=0; **(b)** $t/T$=0.2; **(c)** $t/T$=0.4; **(d)** $t/T$=0.6 **(e)** $t/T$=0.8; and **(f)** $t/T$= 1.0.

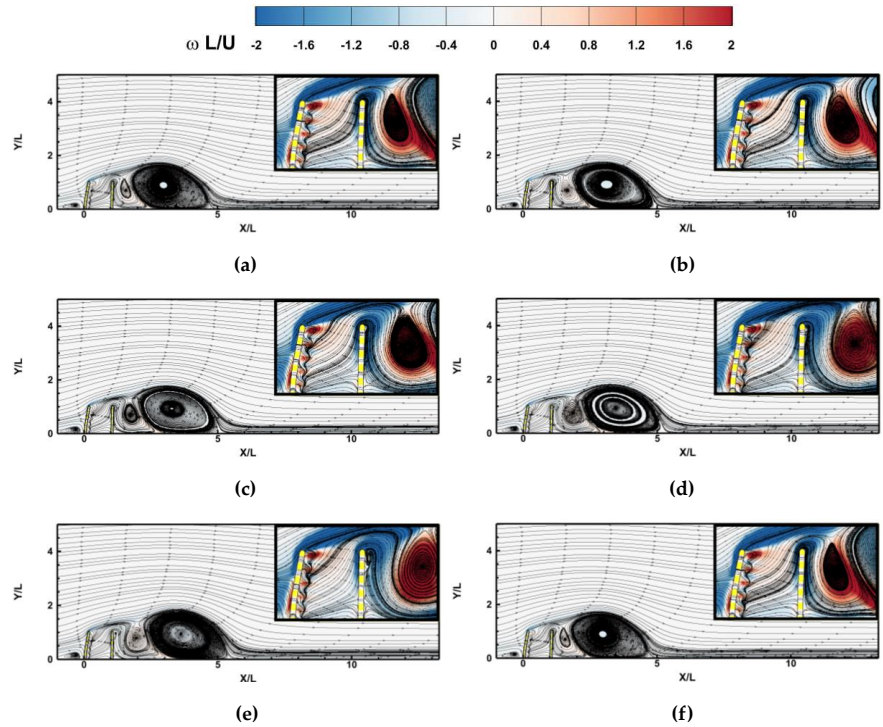

**Figure 30:** Dimensionless vorticity and streamlines plots representing wake structures of a perforated tandem EVG ($\phi$ = 0.25), for $\beta$ = 0.1 at fixed $\gamma$ = 0.5 and Re=400, shown at different instances: (a) t/T=0; (b) t/T=0.2; (c) t/T=0.4; (d) t/T=0.6 (e) t/T=0.8; and (f) t/T= 1.0.

Figures 25–30 illustrate the instantaneous vorticity fields at several instances in a cycle (with time period of T) for non-perforated and perforated tandem EVGs respectively under different parameter conditions. Figures 25 and 26 correspond to the Cavity Oscillation regime at $\gamma$=0.5 for fixed $\beta$=1; Figures 27 and 28 represent the VIV regime at $\gamma$=0.01 for fixed $\beta$=1; and Figures 29 and 30 capture the effect of reduced mass ratio at $\beta$=0.1 for fixed $\gamma$=0.5. Collectively, these cases highlight how bending rigidity, mass ratio, and porosity influence wake dynamics and the overall response of tandem EVGs.

For fixed $\beta$=1, the non-perforated tandem EVGs exhibit a cavity oscillation mode at $\gamma$=0.5 (Figure 25), where large, coherent vortices originate near the tip of the upstream EVG and convect downstream over the cavity region. These vortices travel along a shear layer that arches between the EVGs, impinging periodically on the downstream EVG, hence sustaining a low-pressure recirculating zone. The vortex behavior closely resembles that of an open-cavity flow, with shear-layer instabilities maintaining the oscillatory nature of the wake. In contrast, at $\gamma$=0.01 (Figure 27), both upstream and downstream non-perforated EVGs exhibit vortex-induced vibration (VIV), marked by alternating vortex shedding in an antisymmetric pattern typical of bluff-body wakes. The vortices in this regime are smaller, shed independently from each filament, and do not form a shared recirculating cavity.

When porosity is introduced at $\gamma$=0.5 (Figure 26), the shear layer destabilizes, and vortex formation weakens significantly. Coherent vortices are no longer sustained, and no impingement on the downstream EVG is observed, resulting in a stabilized wake, hence exhibiting static reconfiguration mode. At $\gamma$=0.01 with porosity of $\phi$ = 0.25 (Figure 28), vortex shedding still occurs, though the vortex structures are smaller and less coherent. Bleeding flow leakage through the pores weakens the shed vortices and reduces their convective influence downstream, though VIV remains evident, particularly behind the downstream EVG.

For fixed $\gamma$=0.5, reducing the mass ratio to $\beta$=0.1 (Figure 29) leads to smaller and less energetic vortices in the cavity region compared to $\beta$=1, with continued impingement on the downstream EVG but reduced circulation strength. Introducing porosity at this condition (Figure 30) again suppresses cavity dynamics entirely: vortex development becomes diffuse, and no organized shedding or cavity formation is observed. Overall, these results confirm that cavity oscillation is sustained by strong, coherent vortex structures that travel across the inter-spacing between two EVGs and impact the downstream EVG, while VIV is governed by antisymmetric bluff-body shedding. Porosity disrupts vortex coherence and inhibits shear-layer development, eliminating cavity oscillation mode but allowing VIV to persist in weakened form.

To further substantiate the distinction between VIV and cavity oscillation modes, dimensionless pressure ($C_p$) at several time instances for both the cavity oscillation case ($\gamma$ = 0.5) and the VIV case ($\gamma$ = 0.01) are plotted, all at fixed $\beta$ = 1 and $\phi$ = 0. Figures 31 and 32 illustrate the pressure contours for the cavity oscillation regime for $\phi$ = 0 and $\phi$ = 0.25 respectively at $\gamma$ = 0.5 and $\beta$ = 1. In the non-perforated case (Figure 31), a clear low-pressure recirculating zone is sustained between the filaments, consistent with the coherent vortex formation observed in Figure 25. This low-pressure zone arises due to shear-layer instabilities and periodic vortex impingement on the downstream EVG. When porosity is introduced ($\phi$ = 0.25, Figure 32), this low-pressure region collapses, and no coherent recirculating structure is observed. The pressure field becomes more stabilized, supporting the suppression of cavity oscillation and transition to a static reconfiguration mode, as observed in Figure 26.

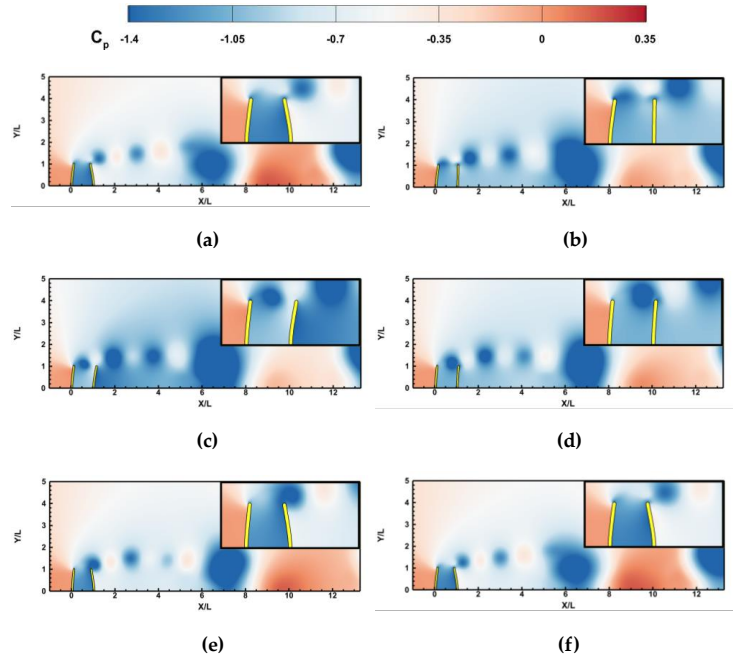

**Figure 31:** Instantaneous dimensionless static pressure contours ($C_p$) for a non-perforated tandem EVG ($\phi = 0$), at $\gamma = 0.5$ for fixed $\beta = 1$ and Re=400, shown at different instances: (a) t/T=0; (b) t/T=0.2; (c) t/T=0.4; (d) t/T=0.6 (e) t/T=0.8; and (f) t/T= 1.0.

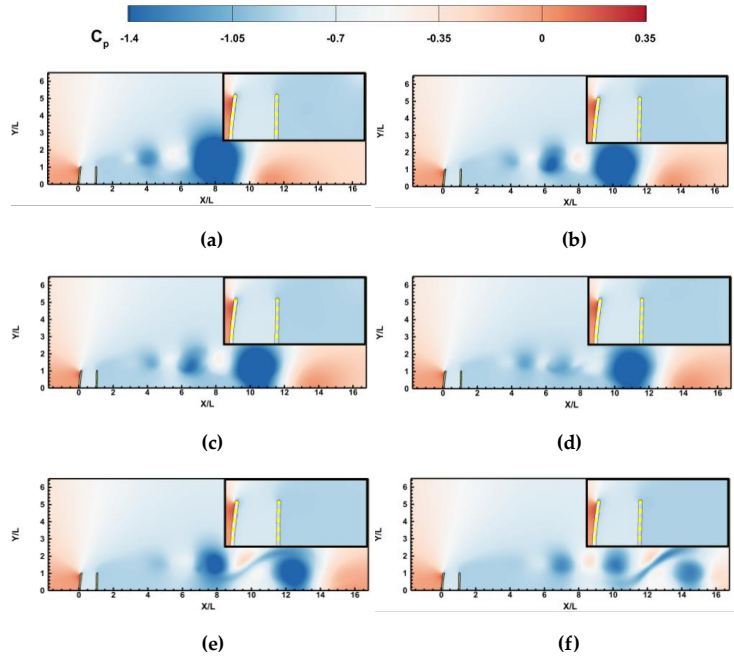

**Figure 32: Instantaneous dimensionless static pressure contours ($C_P$) for a perforated tandem EVG ($\phi = 0.25$), at $\gamma = 0.5$ for fixed $\beta = 1$ and Re=400, shown at different instances: (a) t/T=0; (b) t/T=0.2; (c) t/T=0.4; (d) t/T=0.6 (e) t/T=0.8; and (f) t/T= 1.0.**

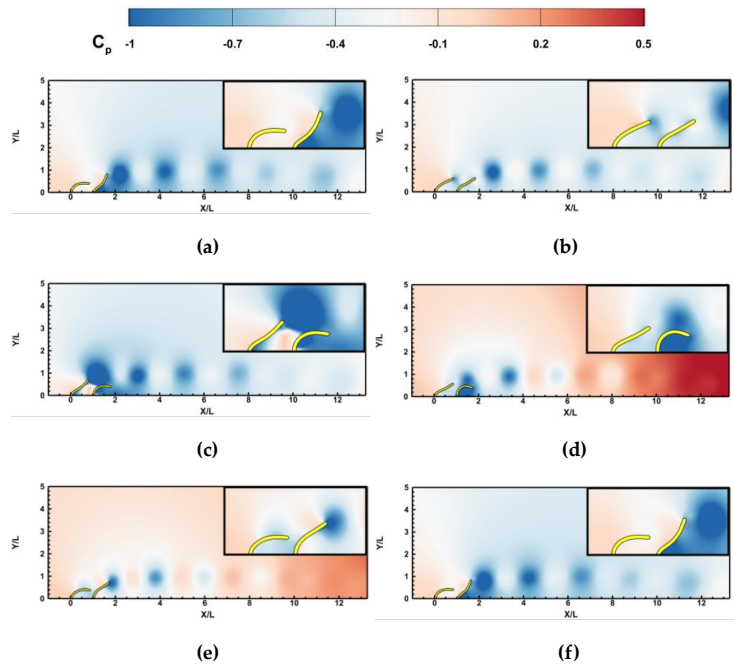

**Figure 33:** Instantaneous dimensionless static pressure contours ($C_P$) for a non-perforated tandem EVG ($\phi = 0$), at $\gamma = 0.01$ for fixed $\beta = 1$ and Re=400, shown at different instances: **(a)** t/T=0; **(b)** t/T=0.2; **(c)** t/T=0.4; **(d)** t/T=0.6 **(e)** t/T=0.8; and **(f)** t/T= 1.0.

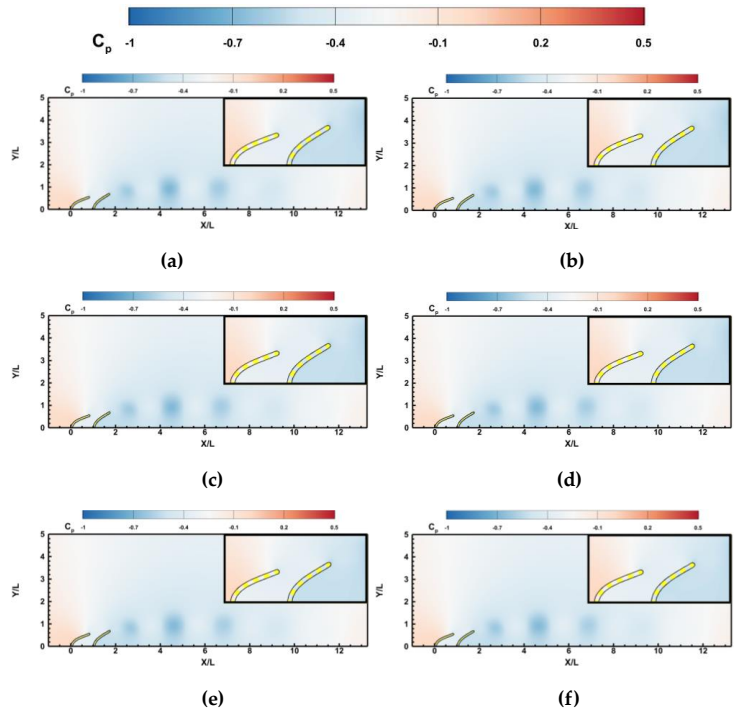

**Figure 34: Instantaneous dimensionless static pressure contours ($C_P$) for a perforated tandem EVG ($\phi$ = 0.25), at $\gamma$ = 0.01 for fixed $\beta$ = 1 and Re=400, shown at different instances: (a) t/T=0; (b) t/T=0.2; (c) t/T=0.4; (d) t/T=0.6 (e) t/T=0.8; and (f) t/T= 1.0.**

Similarly, Figures 33 and 34 correspond to the VIV regime for $\phi$ = 0 and $\phi$ = 0.25 respectively at $\gamma$ = 0.01 and $\beta$ = 1. In both the non-perforated and perforated cases, alternating low-pressure regions consistent with vortex shedding are evident behind each EVG, reaffirming the continued presence of VIV even when porosity is introduced ($\phi$ = 0.25). However, in the perforated case (Figure 34), the pressure distribution is weaker and more diffuse, consistent with the observation that porosity reduces the strength of vortex shedding due to leakage flow and weakened wake interactions.

### 6.8. Lock-in Analysis

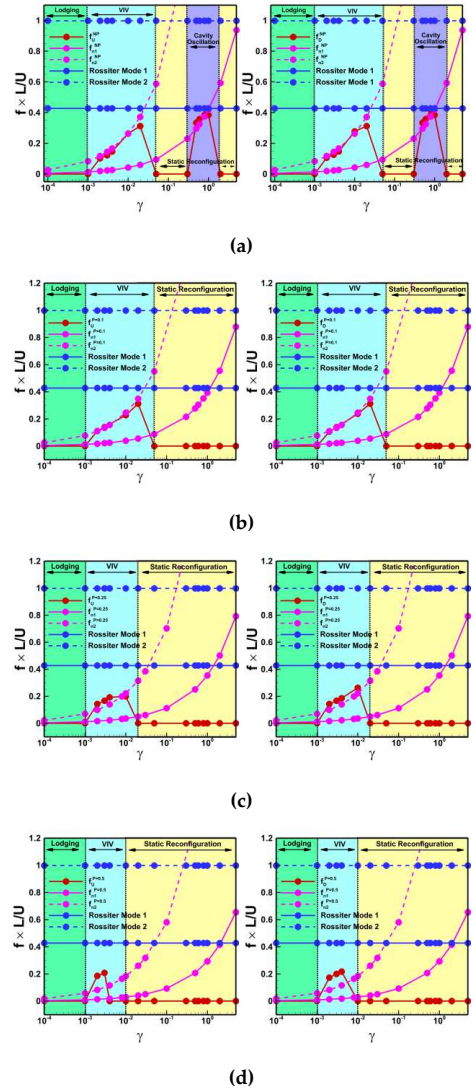

**Figure 35:** Natural frequency of the tandem EVGs for varying porosity levels for β = 1 and Re= 400: (a) φ = 0; (b) φ 0.1; (c) φ = 0.25; and φ = 0.5.

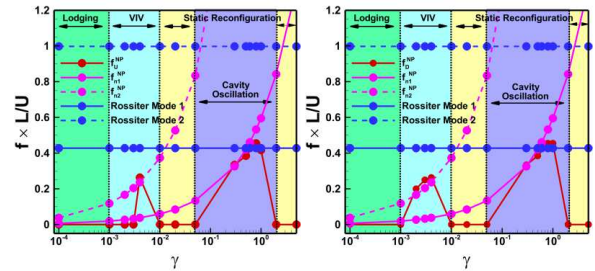

**Figure 36: Natural frequency of the non-perforated tandem EVGs for β = 0.1 and Re=400.**

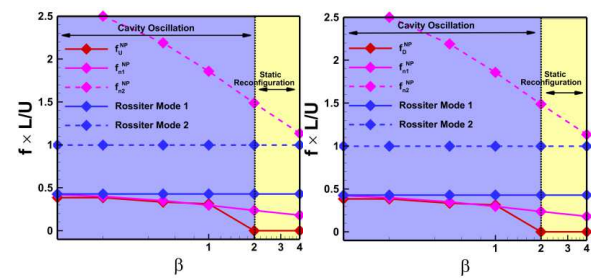

**Figure 37: Natural frequency of the non-perforated tandem EVGs for γ = 0.5 and Re= 400.**

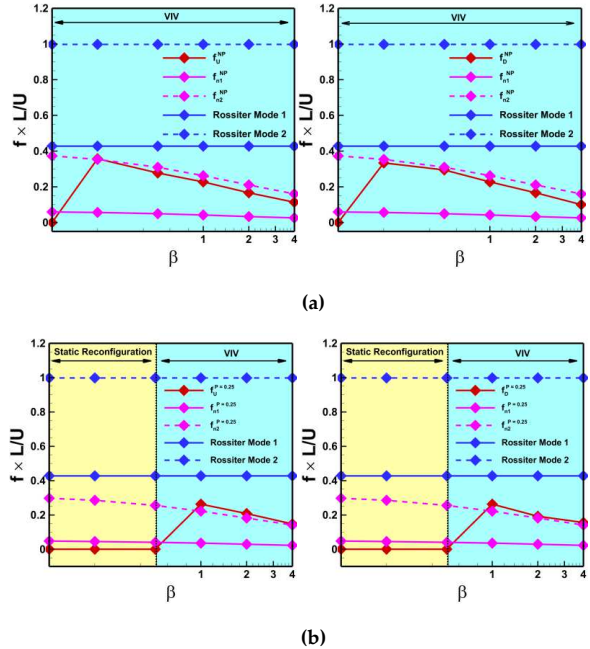

**Figure 38: Natural frequency of the tandem EVGs for varying porosity levels for fixed $\gamma = 0.01$ and Re= 400: (a) $\phi = 0$; and (b) $\phi = 0.25$.**

The oscillatory behavior of tandem EVGs is fundamentally governed by the interaction between their natural frequencies and the unsteady forces generated by vortex shedding. When the vortex shedding frequency synchronizes with one of the structure's natural frequencies, the EVGs undergo a lock-in phenomenon, leading to large-amplitude oscillations. As established in prior theoretical models based on nonlinear Van der Pol oscillators[114], lock-in does not require a perfect one-to-one frequency match; instead, a range of frequency ratios may result in lock-in, with the width of this region depending on the magnitude of the fluctuating loads. Larger fluctuating loads tend to widen the lock-in window, allowing the system to synchronize over a broader range of frequencies. Following the framework developed in our previous work[45], the non-dimensionalized natural frequencies for perforated EVGs (clamped at one end) is given by:

$$f_{ni} = \frac{k_i^2}{2\pi} \sqrt{\frac{\gamma_P}{\beta_P + C_m \frac{\pi}{4}}}$$

(8)

where $k_1 = 1.875$ and $k_2 = 4.694$ are the coefficients of the first and second natural frequencies respectively. In this study the bending rigidity $\gamma$ is kept identical for both perforated and non-perforated EVGs, i.e., $\gamma_P = \gamma_{NP}$, by appropriately adjusting the Young's modulus E in perforated cases using the porosity correction factor K (calculated using Equation 7). $\beta_P$ is the mass ratio for perforated cases and can be calculated using $\beta_P$=X$\beta_{NP}$, where X is the porosity dependent correction factor calculated using $X = \frac{[1-N(B-2)]B}{N+B}$, where, $B = \frac{b}{b+d}$; and N is number of holes.

The added mass coefficient $C_m$ also depends on the perforation and pore layout [89,99,115–124]. However, to estimate the natural frequency, the added mass is assumed $C_m$ =1, similar to the value reported for non-perforated EVGs in the literature[42,45,82], due to the lack of data under conditions directly comparable to this work. For non-perforated EVGs, where the porosity correction factor X=1, Equation (8) simplifies to the equation reported in the literature for non-perforated structures[42,45,82].

Additionally, for cases resembling open-cavity flow behavior (cavity oscillation mode), the dimensionless oscillation frequency can be estimated using Rossiter's semi-empirical formula established in literature[125].

$$f_c = \frac{n-\alpha}{\frac{1}{\kappa}+M} * \frac{1}{d} \qquad (9)$$

where n is the Rossiter mode number (n = 1, 2), $\kappa$=0.57 and $\alpha$=0.25 are empirical constants, and M is the Mach number such that M=0 for incompressible flow conditions.

In the Figures 35-38, the dimensionless oscillation frequency or vortex shedding frequency (shown by solid red line) extracted from both upstream and downstream EVG responses is plotted along with the calculated first natural frequency (shown by solid purple line) and second natural frequency (shown by dashed purple line), and the Rossiter mode frequencies (shown by blue line). These plots enable identification of lock-in phenomena and provide insight into how dynamic regimes transition with changes in bending rigidity, mass ratio, and porosity.

Across all cases, a consistent trend emerges. In the VIV mode, both the upstream and downstream EVGs oscillation frequency always locks onto the second natural frequency ($f_{n2}$) and is far away from the first natural frequency ($f_{n1}$) of the tandem EVG for the considered test conditions. In contrast, the cavity oscillation mode exhibits lock-in with the first natural frequency ($f_{n1}$), and interestingly the oscillation frequency also aligns closely with the first Rossiter mode, reinforcing that cavity oscillation is a unique VIV mode where the vortex shedding frequency aligns with both the natural frequency of the filaments and the Rossiter frequency. VIV behavior reflects bluff-body shedding dynamics, while cavity oscillations are also governed by shear-layer feedback characteristic of open-cavity flows.

The introduction of porosity profoundly modifies the frequency dynamics. As perforation increases, the effective mass ratio of the EVG decreases, leading to an increase in natural frequencies for perforated structures. Consequently, the lock-in and mode transitions shift toward lower γ and higher β values which can provide more energetic structural response. Despite this frequency reduction, VIV persists across all porosities, though at lower frequencies and with smaller oscillation amplitudes. In contrast, the cavity oscillation mode is entirely suppressed when perforations are introduced. The presence of pores disrupts the shear-layer formation and feedback necessary to sustain cavity oscillations, eliminating the cavity mode altogether.

## 7. Conclusion

This study investigated the fluid–structure interaction and dynamic behavior of tandem perforated and non-perforated elastic vortex generators (EVGs) through high-fidelity two-way coupled FSI simulations, focusing on the effects of bending rigidity, mass ratio, and porosity while keeping Reynolds number Re=400 fixed. Three dynamic response modes—Lodging, Vortex-Induced Vibration (VIV), and Static reconfiguration—were observed for both perforated and non-perforated tandem EVGs. However, a distinct Cavity Oscillation mode, locked to the first natural frequency and the first Rossiter mode, emerged only in tandem non-perforated EVGs, highlighting a unique instability mechanism absent in perforated configurations.

Porosity was found to increase the natural frequencies of EVGs, shifting the dynamic modes (VIV regions) and suppressing the cavity oscillation mode entirely. While VIV persisted at different porosity levels, it occurred at lower frequencies and smaller amplitudes in perforated tandem EVGs. Analysis of oscillation amplitudes showed that the downstream EVG consistently experienced larger oscillations compared to the upstream EVG as well as single EVG, due to periodic excitation from the upstream wake. Drag analysis further revealed that the downstream EVG experienced reduced drag relative to the upstream EVG due to wake shielding; however, drag increased with porosity as more flow was transmitted through the upstream EVG. It is also noteworthy that, in the presence of cavity oscillation, the downstream non-perforated EVG experienced negative drag, which is a result of strong suction forces generated by the low-pressure recirculating wake between the EVGs, which pulls the downstream EVG towards the upstream EVG. This effect vanished with perforation, as bleeding flow disrupted cavity formation and restored forward momentum. Local flow studies confirmed that vortex shedding originated at the EVG tips, with perforated EVGs producing smaller, dissipative vortical structures due to bleeding flow. FFT analyses showed that oscillations in VIV modes locked onto the second natural frequency, with higher harmonics significantly damped in perforated tandem cases. Overall, perforation was shown to fundamentally shift dynamic transitions, suppress cavity-driven instabilities, and modulate the wake dynamics without eliminating VIV.

### ACKNOWLEDGMENTS

We are grateful to the Texas Advanced Computing Center (TACC) at The University of Texas at Austin for providing the high-performance computing resources used to carry out the simulations in this study.

### AUTHOR DECLARATIONS

**Conflict of Interest**

The authors have no conflicts to disclose.

### DATA AVAILABILITY

The data that support the findings of this study are available from the corresponding author upon request.

Physics of Fluids

AIP Publishing